\documentclass[useAMS,usenatbib]{mn2e}
\usepackage{natbib}
\usepackage{amsmath}
\usepackage{graphicx}
\usepackage{epstopdf}
\usepackage{aas_macros}
\usepackage{amssymb}

\newcommand{\bvri}{\protect\hbox{$BV\!RI$} }

\newcommand{\dmb}{\protect\hbox{$\Delta m_{15}(B)~$}}

\newcommand{\ion}[2]{#1$\;${\small{#2}}\relax}
\newcommand{\about}{$\sim\!\!$~}
\newcommand{\kms}{km~s$^{-1}$}
\mathchardef\mhyphen="2D
\newcommand{\lcdm}{\protect\hbox{$\Lambda {\rm CDM}~$}}

\title[Dark Energy Constrains from LOSS Sample]{Constraints on Dark Energy with the LOSS SN~Ia Sample}

\author[Ganeshalingam, Li, \& Filippenko]{Mohan Ganeshalingam,$^{1,2}$\thanks{E-mail:
mganeshalingam@lbl.gov} Weidong Li,$^{1,3}$ Alexei~V.~Filippenko$^{1}$ \\
$^{1}$Department of Astronomy, University of California, Berkeley, CA 94720-3411, USA \\
$^{2}$Lawrence Berkeley National Laboratory, Berkeley, CA 94720, USA\\
$^{3}$Deceased December 12, 2011
\\
}

\begin{document}

\date{Accepted  . Received   ; in original form  }

\pagerange{\pageref{firstpage}--\pageref{lastpage}} \pubyear{2012}

\maketitle

\label{firstpage}

\begin{abstract}
We present a cosmological analysis of the Lick Observatory Supernova Search (LOSS) Type Ia supernova (SN~Ia) photometry sample introduced by \cite{ganeshalingam10a}. These supernovae (SNe) provide an effective anchor point to estimate cosmological parameters when combined with data sets at higher redshift. The data presented by \cite{ganeshalingam10a} have been rereduced in the natural system of the KAIT and Nickel telescopes to minimize systematic uncertainties. We have run the light-curve-fitting software SALT2 on our natural-system light curves to measure light-curve parameters for LOSS light curves and available SN~Ia data sets in the literature. We present a Hubble diagram of 586 SNe in the redshift range $z=0.01$--1.4 with a residual scatter of 0.176 mag. Of the 226 low-$z$ SNe~Ia in our sample,  91 objects are from LOSS, including 45 without  previously published distances. Assuming a flat Universe, we find that the best fit for the dark energy equation-of-state parameter $w = -0.86^{+0.13}_{-0.16}$ (stat) $\pm 0.11$ (sys) from SNe alone, consistent with a cosmological constant. Our data prefer a Universe with an accelerating rate of expansion with 99.999\% confidence. When looking at Hubble residuals as a function of host-galaxy morphology, we do not see evidence for a significant trend, although we find a somewhat reduced scatter in Hubble residuals from SNe residing within a projected distance $ < 10$ kpc of the host-galaxy nucleus ($\sigma = 0.156$~ mag). Similar to the results of \cite{blondin11a} and \cite{silverman12c}, we find that Hubble residuals do not correlate with the expansion velocity of \ion{Si}{II} $\lambda 6355$ measured in optical spectra near maximum light. Our data are consistent with no presence of a local ``Hubble bubble." Improvements in cosmological analyses within low-$z$ samples can be achieved by better constraining calibration uncertainties in the zero-points of photometric systems.
\end{abstract}


\section{Introduction}
Type Ia supernovae (SNe~Ia) are believed to be the thermonuclear explosion of a white dwarf undergoing mass transfer from a companion star; see \cite{hillebrandt00a} for a review. At peak brightness, the luminosity of a SN~Ia ($M_{B} \approx -19.2$ mag) can rival that of its host galaxy and is reasonably standard from event to event ($\sigma \approx 0.5$ mag). Observed correlations between light-curve properties and luminosity further allow SNe~Ia to be standardized to within \about\,0.2 mag \citep[\about\,10\% in distance; ][]{phillips93a,riess96b,hamuy96c,tripp98a,phillips99a}. These properties make SNe~Ia an ideal distance indicator on extragalactic scales.

Application of large samples of SNe~Ia led to the discovery that the Universe is currently accelerating in its expansion \citep{riess98a,perlmutter99a}; see \cite{filippenko05} for a review. Subsequent observations of large samples of SNe~Ia out to high redshifts \citep{riess04a,astier06a,riess07a, wood-vasey07a,hicken09b,kessler09b,amanullah10a,sullivan11b,suzuki12a} have led to precise estimates of cosmological parameters when combined with measurements of baryon acoustic oscillations (BAO) and anisotropy in the cosmic microwave background (CMB). Ongoing surveys to collect large samples of SN~Ia light curves include the Palomar Transient Factory \citep[PTF; ][]{rau09a}, the Panoramic Survey Telescope \& Rapid Response System \citep[Pan-STARRS; ][]{kaiser10a}, the Catalina Real-time Transient Survey \citep[CRTS; ][]{drake09a}, the Lick Observatory Supernova Search \citep[][]{filippenko01a}, and the La Silla QUEST survey \citep{hadjiyska12a}, among others.

Reconstructing the expansion history of the Universe with SNe~Ia requires data sets spanning the widest possible range in redshift. Nearby SNe are less sensitive to cosmological parameters, but they provide an anchor on the Hubble diagram to determine the relative brightness of a SN without the influence of cosmological parameters. Leverage on cosmological parameters comes from the luminosity distance to higher redshift SNe sensitive to the details of cosmological evolution. To this end, data sets at both ends of the redshift range play important roles in determining cosmological parameters.

Many groups have undertaken the challenge of collecting large samples of SN~Ia light curves. The first sample of 29 low-$z$ objects was presented by the Cal\'{a}n/Tololo Supernova Survey \citep{hamuy96a}. The SN group at the Harvard/Smithsonian Center for Astrophysics (CfA)  has produced four large samples of low-$z$ SN~Ia light curves \citep{riess99a,jha06b,hicken09a,hicken12a} comprising a significant bulk of published low-$z$ light curves. The Carnegie Supernova Project (CSP) has released two data sets of SN~Ia light curves \citep{contreras10a,stritzinger11a} with an impressive wavelength coverage of $uBgVri$ along with $YJHK$ for a fraction of the objects. Moreover, soon the Nearby Supernova Factory (SNFactory) is expected to release their dataset of spectrophotometric observations that can be used to synthesize measurements in any arbitrary optical broad-band filter \citep{aldering02a}.

At intermediate redshifts ($0.05 < z < 0.35$), the Sloan Digital Sky Survey (SDSS) has published light curves for 146 SNe~Ia in $ugriz$ \citep{holtzman08a} from observations of Stripe 82. At higher redshifts ($0.15 < z < 1.1$), the Supernova Legacy Survey (SNLS) has released the results of their three-year rolling survey \citep{guy10a}, netting a total of 252 objects with $griz$ photometry. ESSENCE (Equation of State: SupErNovae trace Cosmic Expansion) has published their first set of $R$- and $I$-band light curves for 102 objects \citep{miknaitis07a} and should soon publish photometry of their 6-year sample of 228 SNe Ia \citetext{Narayan et al., in prep.}. The search for SNe~Ia extends out to space with programmes utilizing the {\it Hubble Space Telescope (HST)} to find even more-distant SNe to probe the epoch of cosmic deceleration, constrain the evolution of dark energy (DE), and search for possible evolution of SNe~Ia \citep[e.g.,]{riess04a,riess07a,rodney12a,jones13a}. Imminent or future surveys to gather large data sets at high redshift include the Dark Energy Survey (DES) and the Large Synoptic Survey Telescope (LSST).

SN~Ia cosmological analyses utilize an observed correlation between light-curve width and intrinsic luminosity to standardize SN~Ia luminosity \citep[i.e., the ``Phillips relation"; ][]{phillips93a,hamuy96c,phillips99a}. In addition to applying corrections to light-curve width, distance fitters also make assumptions about SN colour and host-galaxy extinction to further standardize SN~Ia luminosity and improve distance estimates \citep{tripp98a,tripp99a}. Many tools have been developed using well-observed SNe as a training set to refine these relationships and produce reliable distance estimates, including Multicolour Light-Curve Shape \citep[MLCS2k2; ][]{riess96b,jha07a}, Spectral Adaptive Light-curve Template \citep[SALT2; ][]{guy07a,guy10a}, SiFTO \citep{conley08a}, and BayeSN \citep{mandel09a,mandel11a}. 

With the success of the light-curve width and colour parameters, the question has turned to what other observables correlate with luminosity (but remain uncorrelated with light-curve width and SN colour) to further improve distance estimates. For example, as discussed in Section 5.3, ejecta velocity may be a significant variable \citep{wang09a,foley11a,foley11b}. Many studies have tried to correlate Hubble residuals with other light-curve \citep{folatelli10a}, spectral \citep{bailey09a,blondin09a,silverman12c}, and host-galaxy properties \citep{kelly10a,sullivan10a,lampeitl10a}. The relative ease of access to small- or medium-aperture telescopes makes low-$z$ samples an ideal choice for pursuing a third parameter.

In this paper, we present a cosmological analysis highlighting the addition of the Lick Observatory Supernova Search (LOSS) sample of nearby SNe~Ia ($z < 0.05$).  LOSS is an ongoing survey to find and monitor optical transients in the nearby Universe within days or weeks of explosion \citep{li00a,filippenko01a,li03b}\footnote{The discovery and automatic filtered photometry of SN 2012cg just a short time after explosion \citep{silverman12d} illustrates some recently implemented modifications to our search.}.  The LOSS SN~Ia sample is the first data release of high-quality \bvri\ light curves of 165 objects observed in the years 1998--2008; see \cite{ganeshalingam10a} for details. The light curves of these objects are well sampled (typical cadence of one epoch every 3--4 days) and on average start a week before maximum light. The LOSS sample has been used to measure the rise-time distribution of SNe~Ia \citep[][]{ganeshalingam11a}, correlate photometric properties with spectral features \citep{silverman12c}, and identify specific subsets of SNe~Ia that may prove to be more precise distance indicators \citep{wang09a,foley11a}.

The aim of this paper is to combine the LOSS dataset with previously published samples to derive constraints on the cosmological parameters $\Omega_{\rm m}$, $\Omega_{\rm DE}$,  and the DE equation-of-state parameter $w = P/(\rho c^{2})$, where $P$ is the pressure and $\rho$ is the density. In addition, we study the effects of systematic uncertainties, and we look for trends underlying residuals in our cosmological fits. The ultimate goal is to find another parameter (be it photometric, spectroscopic, or host-galaxy property) that correlates with luminosity but is uncorrelated with light-curve width and SN colour to further standardize SN~Ia luminosity.

This paper is organized as follows. In Section \ref{s:data}, we introduce the data sets from various photometric samples across a wide range of redshifts. Section \ref{s:dis_fit} describes the methods by which we measure light-curve properties to estimate distances. The results from our sample of 586 SNe~Ia along with joint constraints from other cosmological probes are presented in Section \ref{s:cosmo_results}. Additionally, we explore the impacts of systematic calibration uncertainties on our ability to determine cosmological parameters. In Section \ref{s:phy_sys}, we focus on the low-$z$ sample (a significant fraction from LOSS) to search for physical systematics which may hint at the long-sought third parameter for improving SN~Ia distance estimates.

\section{Data} \label{s:data}
In an idealized SN~Ia cosmological analysis, the full dataset would comprise SN~Ia photometry over the entire range of redshifts taken with a perfectly homogeneous telescope network with well-characterized throughput curves. Currently, the single largest contribution to systematic uncertainties in SN~Ia distance estimates and cosmological parameters is the photometry zero-point \citep{conley11a}. Future data sets from LSST, Pan-STARRS, and DES will bypass this issue by providing a stand-alone, self-consistent dataset with well-understood throughput curves that will replace existing low, intermediate, and high-redshift data sets. However, until the realization of such a dataset, we must rely on combining data sets from multiple observing programs covering the largest possible redshift range. Here we describe the LOSS dataset, as well as the other published data sets used for our cosmological analysis. 

\subsection{The LOSS Sample}
The LOSS sample is the first data release of 165 \bvri\ light curves of nearby (median $z_{\rm CMB} = 0.0194$) SNe~Ia presented by \cite{ganeshalingam10a}. A majority of the light curves in LOSS are well sampled and start a week before maximum light in the $B$ band. The main discovery engine for the SNe~Ia in our sample was the 0.76-m Katzman Automatic Imaging Telescope (KAIT) as part of a targeted search for nearby SNe \citep{li00a,filippenko01a,li03b,leaman11a}. KAIT discoveries account for \about\,70\% of the objects in the LOSS sample. Our photometric follow-up data were acquired with KAIT and the 1-m Nickel telescope at Lick Observatory.

The processing of data is described in detail by \cite{ganeshalingam10a}. In summary, point-spread-function fitting photometry is performed on images from which the host galaxy has been subtracted using templates obtained $> 1$~yr after explosion. Following the convention set by previous photometry data releases \citep{hamuy96a,riess99a,jha06b}, data were transformed to the Landolt system \citep{landolt83a,landolt92a} using averaged colour terms determined over many photometric nights. Calibrations for each SN field are obtained on photometric nights with an average of five calibrations per field.

The motivation to publish light curves on the Landolt system is to allow light curves from multiple sources to be compared easily. However,  the simple colour corrections used to transform natural-system magnitudes to a given standard photometry system is based on coefficients derived from stellar spectral energy distributions (SEDs) which do not accurately describe the SED of a SN (especially at late times when the SN becomes nebular). Applying such colour corrections does not guarantee that the SN photometry will necessarily be on the Landolt system. To properly account for the SN SED, second-order ``S-corrections" \citep{stritzinger02a,wang09b} must be performed using a representative spectral series. In the case of SN~2005cf (a nearby, well-observed SN~Ia caught well before maximum light), S-corrections improved the discrepancy between photometry systems from a root-mean square (rms) of 0.06 mag to 0.03 mag \citep{wang09b}.

However, there is a changing tide in how large data sets of low-$z$ SN~Ia photometry are being released. Starting with the CfA3 dataset \citep{hicken09a} and the CSP data sets \citep{contreras10a,stritzinger11a}, groups have begun to release SN photometry in the natural system of their telescope along with transmission curves of their photometry system, avoiding colour corrections altogether. This has the obvious disadvantage that photometry from two telescope systems cannot be readily compared. Instead, photometry must be transformed from one system to the other using a SN spectral series coupled with the transmission functions of the telescope system.

The true advantage of natural-system photometry is only realized when the transmission function of the telescope system is well characterized. Using natural-system photometry reduces systematic errors in cosmological fits and better constrains cosmological parameters \citep{conley11a,sullivan11b}. The necessary requirement is precise and accurate measurements of the transmission curves to correctly characterize their properties, and low-$z$ SN groups have started to satisfy this requirement \citep{stritzinger11a,hicken12a}. 

The original data release of the LOSS SN~Ia light curves \citep[][]{ganeshalingam10a} was in the Landolt system. To better take advantage of our dataset, here we rerelease our data in the natural system of the KAIT and Nickel telescopes\footnote{At present, data in both the natural and standard photometric systems can be downloaded from http://hercules.berkeley.edu/database .}. We caution that during our follow-up photometry campaign, KAIT went through a change in filter set and three different CCDs with different quantum-efficiency curves. In total, there are four different KAIT transmission curves for each bandpass, and these configurations are referred to as KAIT[1--4]; see \cite{ganeshalingam10a} for details on the characteristics of the bandpasses. Analysis of the dataset as a whole without accounting for differences in the photometry system is ill advised. We recommend either using the results on the Landolt system or transforming all of the data to a common system using an appropriate spectral series \citep[e.g., ][]{hsiao07a}. We document the procedure for transforming our SN photometry into the telescope natural-photometry system in Appendix \ref{app:nat_phot}.

\subsection{Additional Low-Redshift data sets}
In addition to the LOSS dataset, we also include contributions from the Cal\'{a}n/Tololo sample \citep[29 SNe; ][]{hamuy96a}, CfA1 \citep[22 SNe; ][]{riess99a}, CfA2 \citep[44 SNe; ][]{jha06b}, CfA3 \citep[185 SNe; ][] {hicken09a}, CSP \citep[35 SNe; ][]{contreras10a}, and light curves of individual SNe available in the literature. The Cal\'{a}n/Tololo, CfA1, CfA2, and individually published light curves are in the Landolt system, while all of the other samples are presented in the natural system. We will combine all light curves on the Landolt system into the ``Lit" (i.e., literature) sample for the purposes of our analysis. We analyse all other light curves in the natural system of the telescope using published transmission curves.

The common practice within SN cosmology analyses is to combine low-$z$ samples from different observing programmes. There is significant overlap between the CSP, CfA3, and LOSS samples. Instead of combining data for SNe in common, we choose the light curves that have the best temporal coverage and are best sampled.

The overlapping SNe among the data sets offer an opportunity to study systematics between these programmes. In \cite{ganeshalingam10a}, an analysis between the CfA3 and LOSS data sets found that individual SNe could suffer from systematic offsets of up to \about\,0.05 mag, likely owing to differences in calibrations. \cite{ganeshalingam11a} found a typical scatter of 0.03 mag between CSP and LOSS data sets. \cite{hicken12a} found a similar level of disagreement between the three data sets. In Section \ref{ss:sys_unc}, we study the impact that calibration zero-point uncertainties have on our measurement of cosmological parameters.

Throughout the rest of this paper, we refer to the combined sample of LOSS + CfA3 + CSP + Lit as the ``low-$z$" sample.

\subsection{SDSS SNe}
The dataset of 146 SNe~Ia released by SDSS \citep{holtzman08a,kessler09b} fills in the once sparsely populated intermediate-redshift range ($0.1 < z < 0.4$). The data are in the natural system of the SDSS filter set on the AB photometry system \citep{oke74a}. Following \cite{holtzman08a} and \cite{conley11a}, we only use data flagged $< 1024$ and SNe that do not show signs of peculiarity \citep[e.g., SN~2002cx; ][]{li03a,phillips07a}.  We omit $u$- and $z$-band photometry which have been shown to suffer from obvious systematic errors that are not well understood \citep{kessler09b,conley11a}.

\subsection{SNLS SNe}
For higher redshifts ($0.4 < z < 1$), we use the third-year data release from SNLS \citep[SNLS3; ][]{guy10a} consisting of $griz$ photometry of 252 spectroscopically confirmed SNe~Ia in the natural system of the Megacam instrument of the Canada-France-Hawaii telescope (CFHT)\footnote{Downloaded from http://hdl.handle.net/1807/26549 .}. We do not include other older high-$z$ samples such as those of  the Supernova Cosmology Project \citep[SCP; ][]{perlmutter99a} and the High-$z$ SN Search Team \citep{riess98a}, which do not have as strong a handle on systematics and could bias results when combined with the SNLS3 sample. Similarly, we do not include $R$- and $I$-band photometry of 102 SNe from the ESSENCE program \citep{miknaitis07a}. While the ESSENCE sample is large, the systematics are not as well characterized. The SNLS3 sample represents a large sample of high-$z$ SNe with a well-controlled handle on systematic errors \citep{conley11a}.

\subsection{ {\it HST} SNe}
Differentiating different models of DE requires sampling the expansion history at $z > 1$ where optical features have been shifted into the near-infrared. Searching for SNe~Ia at these redshifts is more feasible from a space-based telescope such as {\it HST}. Targeted searches for higher-$z$ SNe have produced a sample in the redshift range $0.2 < z < 1.55$ \citep{riess07a}.

The sample consists of observations using the Advanced Camera for Surveys (ACS) and the Near-Infrared Camera and Multi-Object Spectrograph (NICMOS). Following \cite{conley11a}, we omit SNe at $z < 0.7$ with F606W observations that will include both $U$- and $B$-band features due to the broad wavelength range covered by the filter. To limit the effect of biasing light-curve parameter estimates, we also place the constraint that if there is an early observation at $-20 < t_{B_{\rm max}} < -15$~d, then there must also be an observation at $-8 < t_{B_{\rm max}} < +9$~d. If there is no early-time observation, then we require an observation at $-8< t_{B_{\rm max}} < +5$~d. These requirements minimize the biasing of light-curve parameter estimates \citep{guy10a, conley11a}.


\section{Distance Fitting}\label{s:dis_fit}
The method by which one turns SN~Ia light curves into a distance estimate is a topic of some debate. There is universal agreement regarding a strong correlation between light-curve width and luminosity \citep{phillips93a,phillips99a} that can be used via a linear (or quadratic) correction to standardize SN~Ia luminosity. However, debate continues on how to correct SN luminosity for the effects of variations in intrinsic colour and host-galaxy extinction. In the absence of host-galaxy extinction, SNe~Ia with redder colours are intrinsically fainter \citep{riess96b,jha07a}. Similarly, SNe~Ia that suffer from host-galaxy extinction will also have redder colours and a diminished observed brightness. Both effects act in the same direction (i.e., a redder SN is fainter), making it difficult to disentangle the two effects.

The Bayesian optimist would say, all is not lost: we have prior information on host-galaxy extinction which can only act to make the SN colour redder and is likely drawn from a distribution we can infer \citep[e.g., the galactic line of sight prior; ][]{hatano98a}. Making simple assumptions, we can place constraints on what we think is the probability distribution function of host-galaxy extinction.  If we then train a model on a sample of low-extinction SN light curves, we can apply our model to data light curves to estimate an extinction value within a Bayesian framework. This is the approach taken by the MLCS fitter \citep{riess96b,jha07a}.  BayeSN \citep{mandel09a,mandel11a} takes a similar approach, but also incorporates SN Ia infrared data to infer properties of host-galaxy extinction.

An alternative viewpoint is that the two effects cannot be disentangled and we should deal with the combined effect. The relationship that governs intrinsic colour and luminosity appears to act in the same way that host-galaxy extinction diminishes brightness. Similar to the linear correction between light-curve width and luminosity, another linear correction for the observed SN colour (which includes contributions from both intrinsic colour and host-galaxy extinction) can be applied to further standardize the SN luminosity. This is the approach of two-parameter empirical fits like SALT2 and SiFTO.

MLCS has been successful at predicting SN~Ia distance estimates. However, there are indications that the framework by which it treats host-galaxy reddening may have systematic problems. Using the updated MLCS2k2.v006, \cite{jha07a} found the existence of a somewhat sharp discontinuity in the Hubble expansion rate, suggesting that the region within H$_{0}d_{\rm SN} < 7400$~\kms\ is underdense compared to the global average --- the so-called ``Hubble bubble."  \cite{conley08a} found that this result may be due to how MLCS2k2.v006 treats dust extinction, and using MLCS2k2 with a lower $R_{V}$ value may be more appropriate \citep{hicken09b}.

\cite{kessler09b} found that MLCS2k2.v006 produced distance estimates that favoured a Universe with $w = -0.72 \pm 0.07 ({\rm stat}) \pm 0.11 ({\rm sys})$ assuming a spatially flat Universe (i.e., $\Omega_{\rm m} + \Omega_{\rm DE} = 1$). This is in strong disagreement with other distance fitters, whose results are consistent with $w = -1$, and it was traced back to how MLCS2k2.v006 treats $U$-band observations. The MLCS2k2.v006 $U$-band model was trained on low-$z$ $U$-band observations which are notoriously hard to calibrate. Other distance fitters such as SALT2 were trained on high-$z$ rest-frame $U$-band observations which fall into redder optical bandpasses, typically the $g$ band. Alternatively, the discrepancy might be attributed to intrinsic differences between low-$z$ and high-$z$ SNe~Ia in the ultraviolet rather than to calibrations \citep{ellis08a, cooke11a,foley12c}. However, \cite{conley11a} point out that the direction of this evolutionary difference between low-$z$ and high-$z$ SNe runs in the opposite direction, exacerbating the difference in $U$-band photometry of low-$z$ and high-$z$ SNe. We note that the root of this problem is not the methodology of MLCS2k2, but the implementation of the training process. Current analyses opt to exclude observer-frame $U$-band observations to avoid the issue entirely.

For the purpose of deriving SN~Ia  distances, we will adopt the empirical approach of fitting coefficients to linear corrections for light-curve width and SN colour. We will use SALT2 \citep{guy07a,guy10a} to derive light-curve parameters. Then, using a $\chi^{2}$ minimisation and marginalisation process, we will fit for linear-correction coefficients that minimize the Hubble residuals. Below we discuss our implementation of this procedure.

\subsection{SALT2} \label{ss:salt2}
The first version of SALT  was developed by
SNLS \citep{guy05a}.  SALT2 is an updated version of SALT with a larger training set of light-curve templates; it is the version implemented in this work. SALT2 is
trained on data from low-$z$ SNe from the
literature and SNe from the first two years of the SNLS 
\citep{guy07a,guy10a}. 
 Light-curve parameters are measured using a time-varying spectral series of SNe along with an adopted exponential colour law to model light-curve data given the bandpass of those data. SALT2 measures a parametrisation of the light-curve width ($x_1$), the SN colour ($c$), and the apparent $B$-band magnitude
at maximum light ($m_{B}$). The parameter $x_{1}$ is similar in concept to the stretch parametrisation of light-curve width \citep{perlmutter97a,goldhaber01a}, with increasing $x_{1}$ values corresponding to broader (and more luminous) SNe~Ia. The parameter $c$ is a measurement of the $B_{\rm max}-V_{\rm max}$ pseudocolour relative to the average pseudocolour of the SALT2 training set.

For our implementation of SALT2 on the LOSS data, we do our fitting in the natural-photometry system of LOSS using the KAIT[1--4] and Nickel throughput curves presented by \cite{ganeshalingam10a}.  The LOSS throughput curves are obtained by multiplying the transmission function of each filter by the quantum efficiency of the CCD and the atmospheric transparency at Lick Observatory. The quantum-efficiency curves for the the KAIT and Nickel CCDs are taken from
the manufacturer's claims.  Filter transmission curves for the
two different KAIT filter sets were measured in a laboratory using a
Varian Cary 5000 spectrophotometer. 

Transformations of natural-system magnitudes derived from spectrophotometry of standard stars with our transmission curves from \cite{stritzinger05a} into the Landolt system produced slightly different colour terms than those determined from our multiple nights of photometric observations. Following \cite{stritzinger02a}, we shift our transmission curves in wavelength until we recover our observed colour terms. These shifts can be found in table 5 of \cite{ganeshalingam10a} and are reflected in our transmission curves.

The zero-point adopted in the most recent version of SALT2 is the flux standard BD +17$^\circ$4806; it has well measured Landolt magnitudes as opposed to Vega, which is too bright for most photometric systems and consequently does not have well-measured Landolt magnitudes. Since fitting is done using the transmission curves of the LOSS photometry system, we estimate the magnitudes of BD +17$^\circ$4806 in the the KAIT[1--4] and Nickel systems  by transforming the Landolt magnitudes \citep{landolt07a} using the colour terms in Table 4 of \cite{ganeshalingam10a}.

With light-curve parameters for our full SN~Ia sample obtained, a model for the corrected apparent magnitude of a SN,
$m_{B{\rm ,corr}}$, is adopted which applies linear corrections for the
light-curve width and colour at peak brightness to the measured apparent magnitude,
$m_B$. The corrected apparent magnitude has the form 
\begin{equation}
m_{B{\rm ,corr}} = m_B  + \alpha \times({\rm light\mhyphen curve~width})  - \beta \times({\rm colour}),
\end{equation}
or in terms of the distance modulus ($\mu_{\rm SN}$) and the SALT2 parametrisations of light-curve width ($x_{1}$) and colour ($c$),
\begin{eqnarray}\label{eq:mu}
\mu_{\rm SN} & = & m_{B,{\rm corr}} - M, \\
 & = & m_B + \alpha x_1  - \beta c - M.
\end{eqnarray}
The constants $\alpha$, $\beta$, and $M$ (the fiducial absolute magnitude of a SN~Ia) are determined using a large sample of SNe~Ia fit to minimize residuals against a cosmological model. We marginalize over $M$ in our cosmological fits, while including $\alpha$ and $\beta$ in our fitting process. Marginalizing over  $\alpha$ and $\beta$ has been shown to produce biased results \citep{kowalski08a,conley11a}.

The $\chi^{2}$ statistic of interest given $N$ SNe is 
\begin{equation}\label{eqn:chi2}
\chi^2 =\sum_{s=1}^N \frac{[\mu (z_s,\Omega_{\rm m}, \Omega_{\rm DE},w) -\mu_{{\rm SN},s}]^2}{\sigma_{{\rm m},s}^2 + \sigma_{{\rm pec},s}^2 + \sigma_{\rm int}^2},
\end{equation}
where $\mu (z_s,\Omega_{\rm m}, \Omega_{\rm DE},w) $ is the luminosity distance modulus of the SN in the CMB rest frame given cosmological parameters $(\Omega_{\rm m}, \Omega_{\rm DE},w)$, $\sigma_{\rm m}$ is the measurement 
error in light-curve properties accounting for covariances between measured
parameters, $\sigma_{\rm pec}$ is the uncertainty due to deviations
from Hubble's law induced by gravitational interactions from
neighbouring galaxies, and $\sigma_{\rm int}$ is a constant intrinsic
scatter added to each SN to achieve a reduced $\chi^2 \approx 1$. Uncertainties are assumed to be Gaussian, although there are likely slight departures from this assumption in the light-curve parameter error estimates. We adopt 300~\kms\ as the peculiar velocity for each SN. Changing this value has little effect on the final cosmological results, although increasing $\sigma_{\rm pec}$ slightly decreases the leverage of the lowest-redshift objects by increasing the uncertainty associated with them. For example, for a SN at $z = 0.01 $, increasing the peculiar velocity from 300 to 400~\kms\ increases $\sigma_{\rm pec}$ by 0.066 mag. For a SN at $z = 0.10$, increasing the peculiar velocity increases $\sigma_{\rm pec}$ by 0.007 mag.

Following \cite{conley11a}, we adjust $\sigma_{\rm int}$ for each photometry sample (low-$z$, SDSS, SNLS, and {\it HST}) such that the reduced $\chi^{2}$ for that sample is $\sim 1$ using the best-fitting values for $(\Omega_{\rm m}$, $w)$ in a flat Universe. We fix $\sigma_{\rm int}$ to the values found in Table \ref{t:sig_int} for all subsequent fits. The $\sigma_{\rm int}$ term captures both the intrinsic dispersion in SN~Ia luminosity (after correcting for light-curve width and colour), unaccounted photometric uncertainties within a survey, and possible selection biases. As noted by \cite{conley11a}, this comes at the cost of the ability to discern subtle differences in cosmological models. However, the goal of this paper is to study more common cosmological models, in which case we are willing to accept the increased uncertainty in our distance measurements.

\begin{table}
\caption{Intrinsic dispersion and rms of SN Samples \label{t:sig_int}}
\begin{center}
\begin{tabular}{lcc}
\hline
Sample & $\sigma_{\rm int}$ (mag) & RMS (mag)  \\
\hline 
Low-$z$     &  0.113 & 0.176  \\
SDSS      &  0.097 & 0.160\\
SNLS         &  0.088 & 0.182\\
{\it HST}          &  0.105 & 0.247\\
 \hline
\end{tabular}
\end{center}
\end{table}

\begin{figure}
\begin{center}
\includegraphics[scale=.55]{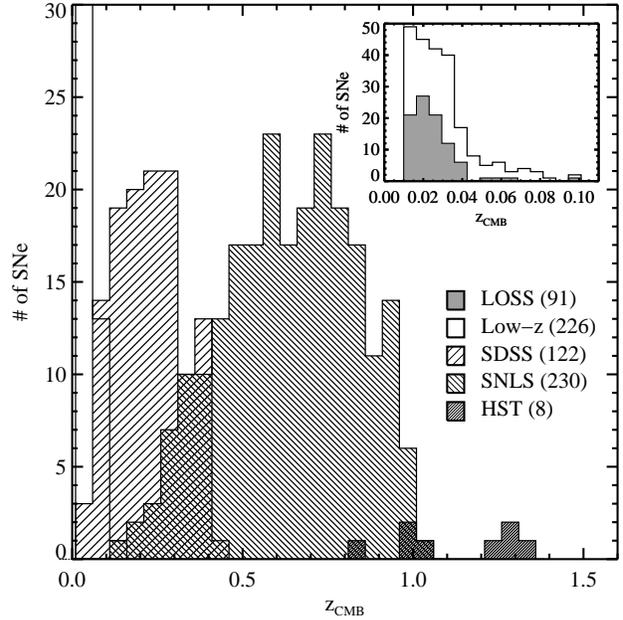}
\end{center}
{\caption[Redshift distribution of photometric samples]{The redshift distribution for the SN sample presented in this work. Each photometric sample is shaded according to the pattern indicated in the legend. Note that the first bin of the low-$z$ sample extends beyond the upper bound of the plot due to the number of objects in that bin. The inset plot shows the redshift distribution for the low-$z$ sample with the LOSS contribution shaded in grey.}\label{figure:z_dis}}
\end{figure}

Cosmological parameters are estimated using two complementary techniques: $\chi^{2}$ minimization and $\chi^{2}$ marginalization. Both techniques were used by \cite{conley11a} and are described in their Appendix B. The $\chi^{2}$ minimization approach uses the  {\tt Minuit} minimization package \citep{james75a} to find the cosmological parameters and linear coefficients that minimize Equation \ref{eqn:chi2}. Parameter errors are estimated using standard assumptions that uncertainties are Gaussian and the model is linear over those uncertainties. The $\chi^{2}$ marginalization procedure calculates Equation \ref{eqn:chi2} over a grid of possible cosmological and linear-coefficient values and then converts $\chi^{2}$ into a relative probability via $P \propto e^{-\chi^{2}/2}$. Confidence intervals are constructed by finding the bounds covering the desired fraction of the total probability. The two different methods will not generally produce exactly the same results since they have different mathematical meanings \citep{upadhye05a, conley11a}. Both codes were downloaded from the SNLS three-year data release website\footnote{Downloaded from \\ https://tspace.library.utoronto.ca/handle/1807/26549 .}.

\begin{figure*}
\begin{center}
\includegraphics[scale=.8]{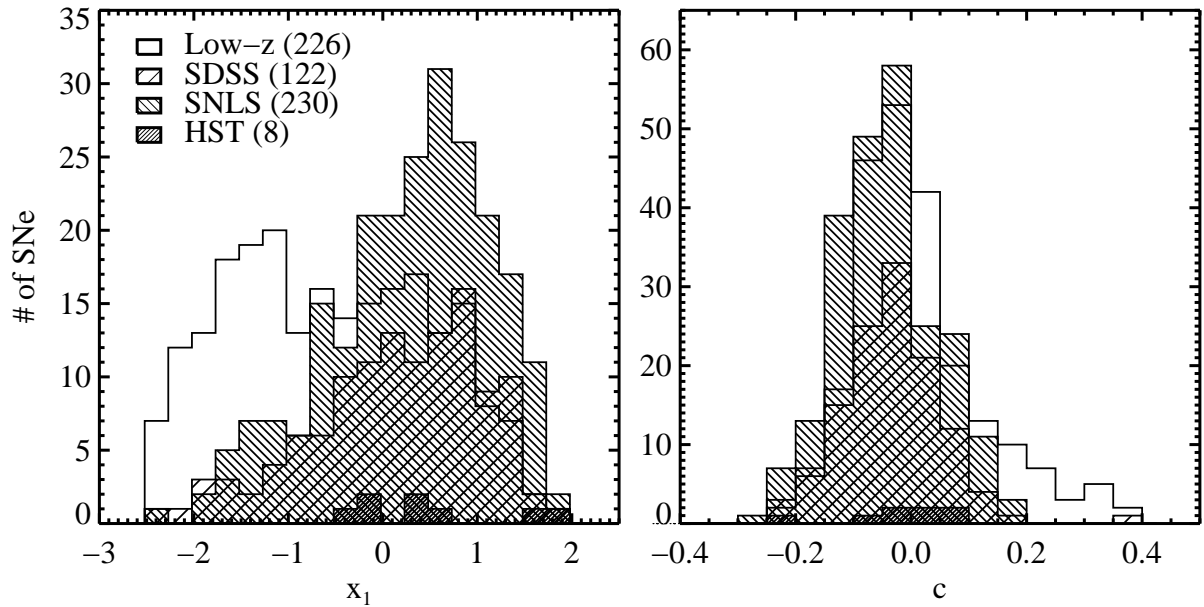}
\end{center}
{\caption[SALT2 parameter distribution for photometric samples]{The SALT2 parameter distribution for the SN sample presented in this work. The left-hand panel shows the $x_{1}$ distribution for our photometric samples and the right-hand panel shows the $c$ distribution. The low-$z$ sample has more underluminous (smaller $x_{1}$) and redder (higher $c$) objects compared to other samples.}\label{figure:x1_c_dis}}
\end{figure*}

\begin{table}
\caption{Summary of SN samples \label{t:samples}}
\begin{center}
\begin{tabular}{lccc}
\hline
Sample & Redshift range & $N_{\rm SN}$   & $ N_{\rm points}^\textrm{a} $ \\
\hline
Low-$z$     &  0.01--0.13  &  226  &  43\\
SDSS      &  0.04--0.42  &  122  &  42\\
SNLS      &  0.12--1.03  &  230  &  37\\
{\it HST}       &  0.84--1.34  &    8  &  13\\
\hline
\multicolumn{4}{c}{{Low-$z$}} \\ \hline             
LOSS      &  0.01--0.10  &   91  &  51\\
CfA3      &  0.01--0.07  &   55  &  32\\
CSP       &  0.01--0.08  &   22  &  77\\
Lit       &  0.01--0.13  &   58  &  31\\
\hline
\multicolumn{4}{l}{$^\textrm{a}$Median number of photometry epochs used in SALT2.}
\end{tabular}
\end{center}
\end{table}

Recent cosmological analyses of SNe~Ia \citep{conley11a,sullivan11b,suzuki12a} have taken into account a correlation found between Hubble residual and host-galaxy mass \citep{kelly10a,sullivan10a,lampeitl10a}. \cite{sullivan10a} found that SNe~Ia in galaxies more massive than $10^{10}~{\rm M_{\odot}}$ are \about\,0.075 mag brighter than SNe hosted in less massive galaxies. \cite{conley11a} implemented this by essentially fitting for two $M$ values\footnote{Note that \cite{conley11a} instead referred to $\mathcal{M}$, which is related to the fiducial absolute mag of a SN~Ia, $M$, by $\mathcal{M} = M - 5 \log_{10} h + 5 \log_{10}{\rm c}$, where c is the speed of light.} depending on a cutoff host-galaxy mass \citep[$10^{10}~{\rm M_{\odot}}$ in the SNLS analysis; ][]{conley11a,sullivan11b}. We are currently in the unfortunate position of not having host-galaxy measurements for a majority of the LOSS objects. We are in the process of combining \bvri\ photometry of our host galaxies taken as part of the LOSS follow-up program with SDSS \citep{abazajian09a} and {\it GALEX} \citep{martin05a} photometry to produce estimates of host-galaxy properties. In this work, we do not make any corrections for host-galaxy mass and instead fit for a single $M$ value.

\subsection{Fitting}
For a given SN, we fit all photometry simultaneously within the range 3000--7000\,\AA, excluding observer-frame $U$-band and $u$-band data (because of the $U$-band anomaly discussed in Section \ref{s:dis_fit}). SN rest-frame $I$, $i$, and $z$ bands are not well represented in the SALT2 training set of SNe; thus, they are also excluded.

\scriptsize
\begin{table*}
\caption{SALT2 Parameters and Distances for SNe \label{t:salt2_param}}
\begin{center}
\begin{tabular}{l c c c c c c c}
\hline
SN & $z_{\rm CMB}$ &  $m_{B}$ (mag)  &  $x_{1}$  &  $c$  & $\mu$ (mag)  &  Sample  & Reference \\
\hline 
SN 1998bp             &   0.010  &  15.294  (0.035)  & -2.520  (0.180)  &   0.197    (0.033)  &  33.363    (0.261)  &  Low-$z$   &  1  \\
SN 2002dj             &   0.010  &  13.880  (0.028)  & -0.276  (0.075)  &   0.004    (0.026)  &  32.889    (0.249)  &  Low-$z$   &  2  \\
SN 2002cr             &   0.010  &  14.178  (0.020)  & -0.601  (0.040)  &  -0.047    (0.019)  &  33.299    (0.246)  &  Low-$z$   &  2  \\
SN 1999cp             &   0.010  &  13.927  (0.027)  & -0.084  (0.043)  &  -0.045    (0.026)  &  33.119    (0.249)  &  Low-$z$   &  2  \\
SN 2002dp             &   0.010  &  14.456  (0.022)  & -0.830  (0.047)  &   0.020    (0.021)  &  33.333    (0.246)  &  Low-$z$   &  2  \\
SN 2006bh             &   0.011  &  14.344  (0.023)  & -1.561  (0.047)  &  -0.063    (0.024)  &  33.377    (0.247)  &  Low-$z$   &  3  \\
SN 1999ee             &   0.011  &  14.872  (0.027)  &  0.719  (0.041)  &   0.237    (0.025)  &  33.289    (0.246)  &  Low-$z$   &  4  \\
SN 2001fh             &   0.012  &  14.062  (0.039)  & -2.412  (0.353)  &  -0.230    (0.035)  &  33.498    (0.234)  &  Low-$z$   &  2  \\
SN 2005bc             &   0.013  &  16.236  (0.022)  & -1.652  (0.117)  &   0.348    (0.022)  &  33.954    (0.216)  &  Low-$z$   &  2  \\
SN 1999ej             &   0.013  &  15.378  (0.029)  & -1.723  (0.087)  &  -0.018    (0.028)  &  34.244    (0.219)  &  Low-$z$   &  2  \\
SN 2001ep             &   0.013  &  14.853  (0.028)  & -0.855  (0.058)  &   0.018    (0.026)  &  33.732    (0.215)  &  Low-$z$   &  2  \\
SN 2006lf             &   0.013  &  13.721  (0.031)  & -1.304  (0.072)  &  -0.211    (0.028)  &  33.260    (0.217)  &  Low-$z$   &  5  \\
SN 2002ha             &   0.013  &  14.679  (0.021)  & -1.373  (0.055)  &  -0.086    (0.020)  &  33.813    (0.211)  &  Low-$z$   &  2  \\
SN 2005al             &   0.013  &  14.848  (0.022)  & -1.174  (0.044)  &  -0.093    (0.024)  &  34.033    (0.212)  &  Low-$z$   &  3  \\
SN 1997E              &   0.014  &  15.097  (0.030)  & -1.612  (0.139)  &   0.014    (0.029)  &  33.877    (0.214)  &  Low-$z$   &  1  \\
SN 1999dq             &   0.014  &  14.397  (0.028)  &  0.836  (0.051)  &   0.045    (0.026)  &  33.438    (0.210)  &  Low-$z$   &  1  \\
SN 1991ag             &   0.014  &  14.448  (0.041)  &  0.697  (0.136)  &  -0.042    (0.031)  &  33.744    (0.212)  &  Low-$z$   &  6  \\
SN 2005kc             &   0.014  &  15.493  (0.022)  & -0.670  (0.057)  &   0.161    (0.024)  &  33.947    (0.206)  &  Low-$z$   &  3  \\
SN 1999dk             &   0.014  &  14.783  (0.029)  & -0.171  (0.069)  &   0.001    (0.026)  &  33.816    (0.206)  &  Low-$z$   &  2  \\
SN 1992al             &   0.014  &  14.461  (0.030)  & -0.226  (0.084)  &  -0.111    (0.027)  &  33.840    (0.206)  &  Low-$z$   &  6  \\
SN 2006N              &   0.014  &  15.091  (0.031)  & -1.939  (0.104)  &  -0.041    (0.029)  &  33.998    (0.206)  &  Low-$z$   &  5  \\
SN 2001bt             &   0.014  &  15.271  (0.029)  & -0.874  (0.070)  &   0.157    (0.027)  &  33.707    (0.203)  &  Low-$z$   &  7  \\
SN 2001fe             &   0.014  &  14.658  (0.030)  &  0.704  (0.129)  &  -0.053    (0.028)  &  33.991    (0.206)  &  Low-$z$   &  5  \\
SN 2000dm             &   0.015  &  15.029  (0.031)  & -1.853  (0.124)  &  -0.058    (0.030)  &  34.003    (0.206)  &  Low-$z$   &  2  \\
SN 2004ey             &   0.015  &  14.704  (0.021)  &  0.008  (0.027)  &  -0.131    (0.022)  &  34.181    (0.199)  &  Low-$z$   &  3  \\
\hline 
\multicolumn{8}{p{6in}}{Only a portion of this table is shown here for guidance regarding its form and content. Uncertainties are 1$\sigma$.}\\
\multicolumn{8}{p{6in}}{{\bf References.} (1) \cite{jha06b}; (2) \cite{ganeshalingam10a}; (3) \cite{contreras10a}; (4) \cite{stritzinger02a}; (5) \cite{hicken09a}; (6) \cite{hamuy96a}; (7) \cite{krisciunas04b}; (8) \cite{riess99a}; (9) \cite{krisciunas04a}; (10) \cite{krisciunas01a}; (11) \cite{holtzman08a}; (12) \cite{guy10a}; (13) \cite{riess07a}.}
\end{tabular}
\end{center}
\end{table*}
\normalsize

\subsection{Selection Criteria}
We employ selection criteria to ensure that we are only including SNe~Ia that are adequately fit by the SALT2 model. We require SALT2 fits to have a reduced $\chi^{2} < 2$, although each fit is also visually inspected. Many of the low-$z$ objects have spectroscopic identifications from \cite{silverman12a} or \cite{blondin12a}, allowing us to securely eliminate peculiar SNe that are not represented in the SALT2 model training set. These include objects similar to SN~2000cx \citep{li01b}, SN~2001ay \citep{krisciunas11a}, SN~2002cx \citep{li03a}, SN~2002es \citep{ganeshalingam12a}, SN~2006bt \citep{foley10b}, or SN~2009dc \citep{yamanaka09a,silverman11a,taubenberger11a}.
 
Cuts on $c$ and $x_{1}$ are required to ensure that we are within a parameter space described by the SALT2 training set. This also helps eliminate peculiar objects that are either subluminous \citep[e.g., SN~1991bg-like objects; ][]{filippenko92a,leibundgut93a,taubenberger08a}, super-Chandrasekhar-mass candidates \citep[e.g., SNLS-03D3bb, SN~2006gz, SN~2007if, SN~2009dc;][]{howell06a,hicken07a,yamanaka09a,scalzo10a,silverman11a,taubenberger11a}, or highly reddened objects that may follow a different host-galaxy extinction law \citep[e.g., SN 2006X; ][]{wang08a}. We restrict our sample to objects with $c < 0.50$ and $-3 < x_{1}< 2$. In general, we found that relaxing these constraints had little effect on our final results, although it significantly increased the number of individual outliers in our fits.

A minimum redshift cutoff is employed to minimize the impact of peculiar velocities. Previous analyses have used cutoffs in the range of $z= 0.01$--0.02. More conservative cutoffs have been used to eliminate any influence of a Hubble bubble. Our analysis does not find evidence for a Hubble bubble, consistent with what is found by \cite{conley07a} and \cite{hicken09b} (see Section \ref{ss:hubble_bubble}). To maximize the number of objects in the low-$z$ sample we adopt a cutoff of $z=0.01$. Increasing our minimum redshift does not significantly alter our results. Objects along a line of sight with a Milky Way reddening of $E(B-V) >  0.20$~mag as measured by the dust maps of \cite{schlegel98a} are also excluded due to concerns that $R_{V} \neq 3.1$ \citep{conley11a}.

The number of objects contributing to our final cosmology sample, the redshift range, and the typical number of points used in our SALT2 fit can be found in Table \ref{t:samples}. Figure \ref{figure:z_dis} shows the redshift distribution for each of the samples in our final cosmology set totaling 586 objects. The inset to Figure \ref{figure:z_dis} reflects the LOSS contribution of 91 out of 226 objects (45 without previously published distances) to the low-$z$ SNe used in this analysis.

In Figure \ref{figure:x1_c_dis}, we plot the $x_{1}$ (left-hand panel) and $c$ (right-hand panel) distribution for our different samples. The low-$z$ sample has the clearest differences compared to other higher redshift samples. This is not unexpected since most low-$z$ SN discoveries are from targeted searches like LOSS or by amateur astronomers which are biased compared to an untargeted sample. SDSS and SNLS, on the other hand, are rolling searches which do not target preselected galaxies. The low-$z$ sample has significantly more underluminous objects (i.e., smaller $x_{1}$) and redder objects (i.e., larger $c$) compared to the other samples.

Table \ref{t:salt2_param} shows our SALT2 light-curve parameter values as well as distance estimates using fitting coefficients derived from our best-fitting cosmology (see Section \ref{s:cosmo_results} for details).

\begin{figure*}
\begin{center}
\includegraphics[scale = .75]{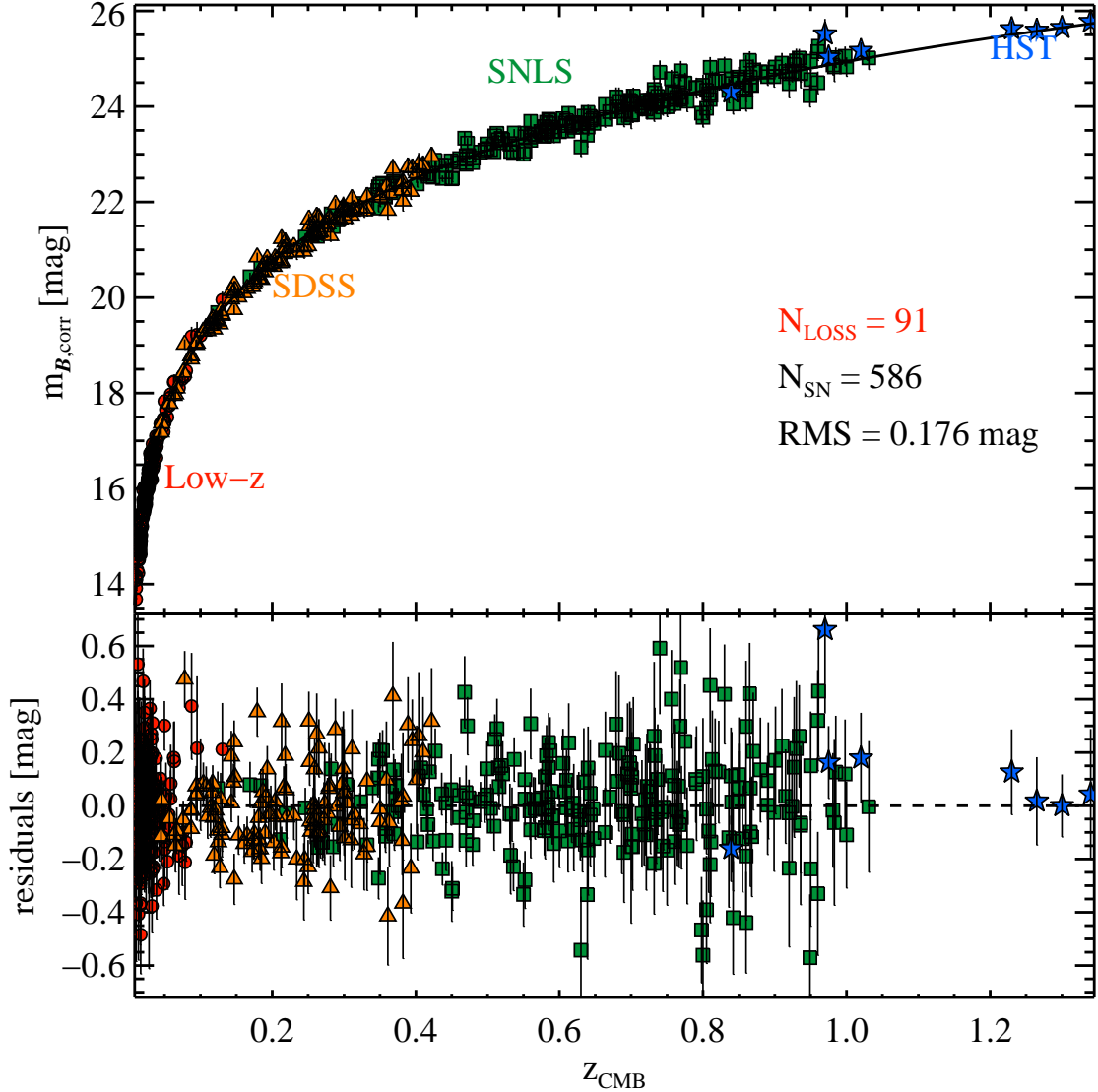}
\end{center}
{\caption[Hubble diagram for 586 SNe~Ia]{Hubble diagram for 586 SNe~Ia. Each photometric sample is coded by the indicated colour. Overplotted as a solid line is the best-fitting cosmology. In the bottom panel we show the residuals of the fit.  Residuals are measured as $\mu_{\rm SN} - \mu(z,\Omega_{\rm m}, \Omega_{\rm DE},w)$.}\label{figure:hubble}}
\end{figure*}


\section{Cosmological Results and Systematics} \label{s:cosmo_results}
In this section, we present the results from our cosmological analysis using the full cosmological sample, with the inclusion of the LOSS sample, which contributes 91 SNe, 45 having distance measurements published for the first time. We present both the results of our $\chi^{2}$ minimisation and marginalisation procedures (see Section \ref{ss:salt2}) and discuss the significance of our results. We then examine the impact of systematic uncertainties and their effect on our ability to measure cosmological parameters.

\subsection{Best-fitting Cosmology}
In Figure \ref{figure:hubble}, we show our combined best-fitting Hubble diagram for the entire cosmological set of 586 SNe. The sample covers the redshift range 0.01--1.34. Overplotted is the  best-fitting cosmology assuming a cosmological constant (i.e., $w = -1$). We find a scatter of 0.176 mag for the entire sample. Objects are colour coded by the sample to which they belong. Plotted in the bottom panel are the residuals as a function of redshift.  We define the Hubble residual as $\mu_{\rm SN} - \mu(z,\Omega_{\rm m}, \Omega_{\rm DE},w)$. SNe that are too bright for their luminosity distance will produce a negative residual. No significant trend is found as a function of redshift (see Section \ref{ss:tension} for further discussion on tensions between data sets).

We measure cosmological parameters in the context of two scenarios. First, we consider a flat Universe dominated by matter and DE. We fit for $\Omega_{\rm m}$ and the DE equation-of-state parameter, $w$. Second, we relax the assumption of flatness and consider a Universe with matter and a cosmological constant, $\Lambda$. We fit for  $\Omega_{\rm m}$ and $\Omega_{\Lambda}$. We present the results of both scenarios in Tables \ref{t:cosmo_flat_results} and \ref{t:cosmo_lambda_results}. Results for both $\chi^{2}$ minimisation and marginalisation are shown, and they are consistent within $1\sigma$ uncertainties. In the results that follow, we present our $\chi^{2}$ marginalisation results which produce more reliable uncertainty estimates.

Assuming a flat Universe dominated by matter and DE, we find using only SN~Ia data that $\Omega_{\rm m} =0.159^{+0.077}_{-0.086} $ and $w=-0.855^{+0.125}_{-0.164}$. We are able to reject a Universe without DE at the $> 4\sigma$ level (99.999\% confidence) using the SN~Ia dataset. Including BAO measurements from \cite{percival10a} and measurements of anisotropy in the CMB from \cite{komatsu11a}, we find significantly tighter constraints of $\Omega_{\rm m} = 0.270^{+0.018}_{-0.012} $ and $w = -1.067^{+0.050}_{-0.046}$. Our data are consistent with the accelerating expansion of the Universe driven by a cosmological constant. This is in excellent agreement with recent results from \cite{conley11a} and \cite{suzuki12a}.

\begin{figure}
\begin{center}
\includegraphics[scale=.55]{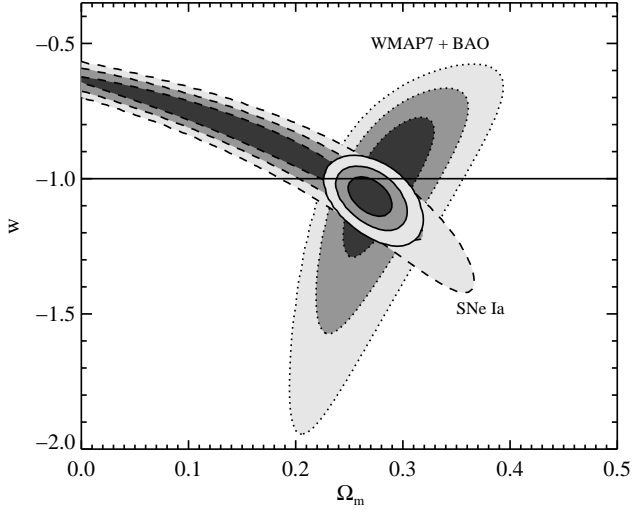}
\end{center}
{\caption[Probability contours for $\Omega_{\rm m}$ and $w$]{Probability contours for $\Omega_{\rm m}$ and $w$ assuming a flat Universe.  Contours represent 68.3\%, 95.4\%, and 99.7\% confidence levels. Constraints from SNe~Ia  and  WMAP7 + BAO are labeled accordingly. The combined constraints lie in the intersection of the two sets of confidence intervals. Overplotted as a solid line is the expected equation-of-state parameter for a Universe with a cosmological constant. Our combined SN~Ia + WMAP7 + BAO results are consistent with the concordance \lcdm cosmology.}\label{figure:cosmo_contour}}
\end{figure}

\begin{table}
\caption{Best-fitting cosmological parameters, flat Universe. \label{t:cosmo_flat_results}}
\begin{center}
\begin{tabular}{lcc}
\hline
& $\chi^{2}$   & $\chi^{2}$ \\
& Minimization  & Marginalisation \\
\hline 
\multicolumn{3}{c}{{SNe only}} \\ \hline
$\Omega_{{\rm m}}$ & $0.151^{ +0.087}_{ -0.117}    $   & $  0.159^{+0.077}_{-0.086}$\\
$w$    &  $-0.813^{+0.155}_{ -0.170} $       & $ -0.855^{+0.125}_{-0.164}$ \\
$\alpha$ &$0.146 \pm 0.007 $&$ 0.147^{+0.010}_{-0.006}$\\
$\beta$ &$3.168^{+0.079}_{-0.078} $ &$ 3.2288^{+0.088}_{-0.070}$ \\
\hline \\[-1.8ex]
\multicolumn{3}{c}{{SNe + CMB + BAO}} \\ \hline
$\Omega_{{\rm m}}$   &  $ 0.270^{+0.014}_{-0.013}$    &  $ 0.270^{+0.018}_{-0.012}$     \\
$w$      & $ -1.055^{+0.047}_{ -0.050}  $  &   $ -1.067^{+0.050}_{-0.046}$  \\
 \hline
\end{tabular}
\end{center}
\end{table}

\begin{table}
\caption{Best-fitting cosmological parameters if $w =-1$. \label{t:cosmo_lambda_results}}
\begin{center}
\begin{tabular}{lcc}
\hline

& $\chi^{2}$   & $\chi^{2}$ \\
& Minimization  & Marginalisation \\
\hline 
\multicolumn{3}{c}{{SNe only}} \\ \hline
$\Omega_{{\rm m}}$ & $0.163^{+0.079}_{ -0.083}$   & $   0.161^{+0.079}_{-0.074}$\\
$\Omega_{\rm \Lambda}$    &  $0.635^{+0.121}_{ -0.127}$       & $ 0.641^{+0.117}_{-0.113} $\\
$\alpha$ &$0.146 \pm 0.007 $&$ 0.147^{+0.010}_{-0.006}$\\
$\beta$ &$3.169^{+0.079}_{-0.078} $ &$ 3.229^{+0.088}_{-0.070} $\\
\hline \\[-1.8ex]
\multicolumn{3}{c}{{SN + CMB + BAO}} \\ \hline
$\Omega_{{\rm m}}$   &  $ 0.265^{+0.013}_{ -0.012}$    & $ 0.264^{+0.019}_{-0.011} $  \\
$\Omega_{\Lambda}$      & $ 0.737^{+0.012}_{-0.013} $  &   $0.738^{+0.015}_{-0.011} $ \\
 \hline
\end{tabular}
\end{center}
\end{table}

\begin{figure}
\begin{center}
\includegraphics[scale=.55]{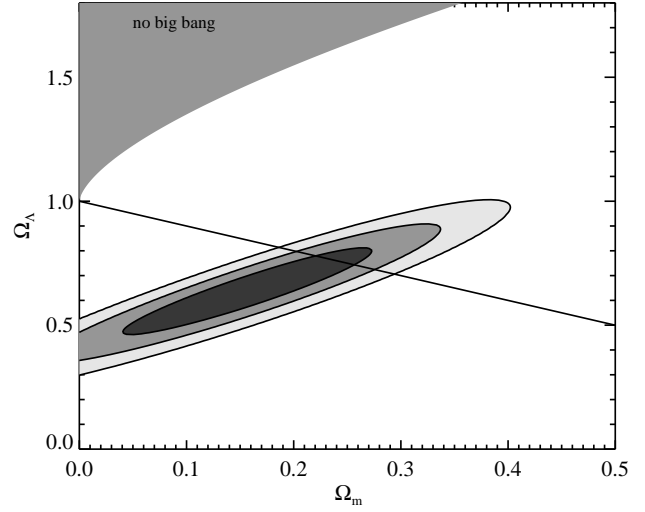}
\end{center}
{\caption[Probability contours for $\Omega_{\rm m}$ and $\Omega_{\Lambda}$ from SNe~Ia]{Probability contours for $\Omega_{\rm m}$ and $\Omega_{\Lambda}$ from SNe~Ia. Overplotted as a solid line is the expectation for a flat Universe.}\label{figure:cosmo_contour2}}
\end{figure}

\begin{figure}
\begin{center}
\includegraphics[scale=.55]{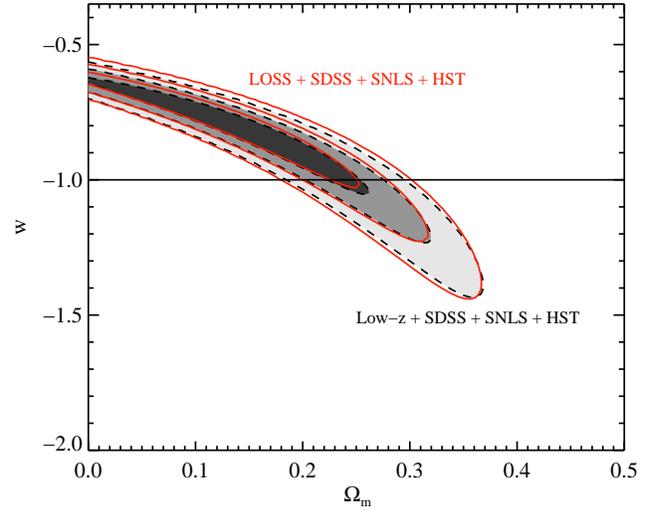}
\end{center}
{\caption[Comparison of Probability contours for $\Omega_{\rm m}$ and $w$]{Comparison of probability contours for $\Omega_{\rm m}$ and $w$ assuming a flat Universe using the entire low-$z$ sample with other samples versus using LOSS with other samples. Constraints from using the entire low-$z$ sample are plotted in grey-scale while constraints from LOSS are outlined in red.}\label{figure:cosmo_contour_comp}}
\end{figure}

In Figure \ref{figure:cosmo_contour}, we show two-dimensional probability contours in the $\Omega_{\rm m}$--$w$ plane for SNe~Ia in blue, WMAP7 + BAO in green, and the joint constraints from SNe~Ia + WMAP7 + BAO in grey. Contours indicate the 68.3\%, 95.4\%, and 99.7\% confidence levels. Overplotted as a black solid line is the expectation from a cosmological constant. When combined with the BAO + CMB data, the SNe offer significantly tightened constraints on cosmological parameters, as expected. Our results are within the 1$\sigma$ confidence interval of $w = -1$.

Next, we allow for a Universe with curvature and DE powered by a cosmological constant. From our SN fits, we find  $\Omega_{\rm m} = 0.161^{+0.079}_{-0.074}$ and $\Omega_{\Lambda} =0.641^{+0.117}_{-0.113}$. When combined with constraints from WMAP7 + BAO, we recover concordance cosmology values of  $\Omega_{\rm m} = 0.264^{+0.019}_{-0.011}$ and $\Omega_{\Lambda} = 0.738^{+0.015}_{-0.011}$.

In Figure \ref{figure:cosmo_contour2}, we show probability contours in the $\Omega_{\rm m}$--$\Omega_{\Lambda}$ plane from only SNe~Ia. Our results are consistent with a flat, $\Lambda$-dominated Universe as indicated by the solid black line.

If we restrict our low-$z$ sample to LOSS data with good fits (104 SNe) in addition to SDSS + SNLS + {\it HST}, we lose only a small degree of statistical leverage. From our fits with just SNe~Ia, we find $\Omega_{m} =  0.151^{+0.083}_{-0.087}$ and $w = -0.835^{+0.125}_{-0.178}$. The only slight increase in statistical error is a strong indication that we are reaching the statistical limit achievable by current low-$z$ data sets. Future improvements will be gained by eliminating systematic errors.

In Figure \ref{figure:cosmo_contour_comp}, we compare the probability contours in the $\Omega_{\rm m}$--$w$ plane using our entire low-$z$ sample to those using only LOSS. The blue shaded contours represent constraints from using all of our low-$z$ + SDSS + SNLS + {\it HST} data. The red outline indicates constraints from LOSS + SDSS + SNLS + {\it HST} data. The contours from using just LOSS are slightly larger than with the entire low-$z$ sample. Comparing the area in the innermost contours, the LOSS  contour is \about\,18\% larger than the low-$z$ contour.

\begin{figure*}
\begin{center}
\includegraphics[scale = .8]{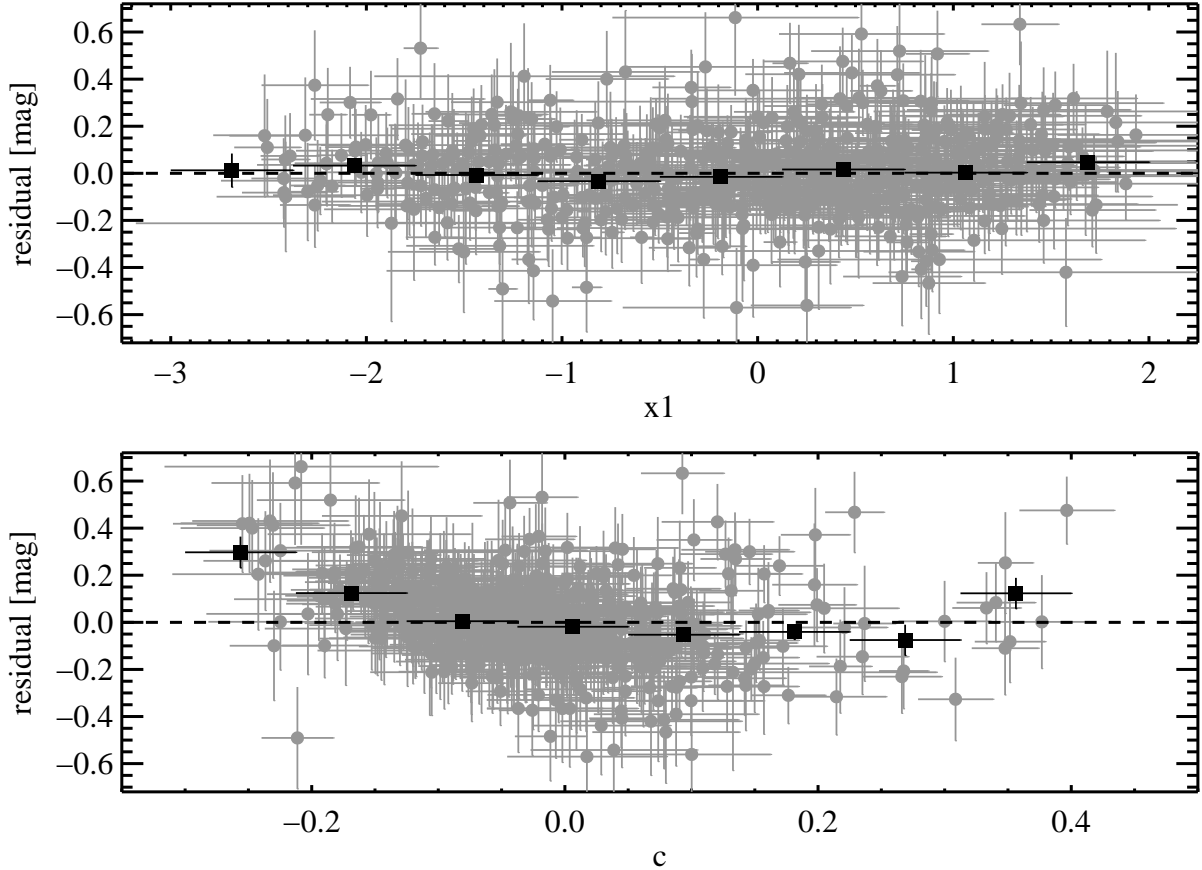}
\end{center}
{\caption[Residuals as a function of SALT2 parameter]{Residual as a function of SALT2 parameter. Residuals are measured as $\mu_{\rm SN} - \mu(z,\Omega_{\rm m}, \Omega_{\rm DE},w)$. We plot the binned error-weighted mean residual as a black square. The bins are of equal width with each point centred at the middle of the bin. In the top panel, we show the Hubble residual against $x_{1}$ where we see no trend. In the bottom panel, we show Hubble residual against $c$. We find a linear trend for $c < - 0.05$, indicative that bluer SNe prefer a lower $\beta$ value.}\label{figure:param_v_res}}
\end{figure*}

\subsection{Tension between data sets}\label{ss:tension}
As a measure of systematic differences between the data sets, we calculate the error-weighted mean residual and its uncertainty in Table \ref{t:tension} for the samples used in this study. We fix $\alpha$ and $\beta$ to their best-fitting values. Overall, we do not see any significant residuals between the data sets. The SNLS dataset shows a slightly higher mean residual, but not at a significant level (\about\,2$\sigma$). The slight trend with increasing redshift is not statistically significant.

Within the low-$z$ data sets, the Lit sample shows the largest mean residual of $-0.062 \pm 0.023$ mag at the 2.7$\sigma$ level. The LOSS, CfA3, and CSP samples, all of which are fit in their natural system, do not show significant mean offsets.

\begin{table}
\caption{Tension between SN samples \label{t:tension}}
\begin{center}
\begin{tabular}{lcc}
\hline
Sample & Weighted mean & Uncertainty  \\
& (mag) & (mag)  \\
\hline
Low-$z$     &  $-$0.005  &  0.012 \\
SDSS      &  $-$0.011  &  0.015 \\
SNLS      &  0.023   &  0.010 \\
{\it HST}       &  0.035   &  0.073 \\
\hline
\multicolumn{3}{c}{{Low-$z$}} \\ \hline             
LOSS      &  0.024   &  0.016 \\
CfA3      &  0.035   &  0.022 \\
CSP       &  $-$0.007  &  0.028 \\
Lit       &  $-$0.062  &  0.023 \\
\hline
\end{tabular}
\end{center}
\end{table}

\subsection{Residuals with SALT2 Parameters} \label{ss:res}
In Figure \ref{figure:param_v_res}, we plot the Hubble residual as a function of $x_{1}$ (top panel) and $c$ (the bottom panel). Recall that we define the residual as $\mu_{\rm SN} - \mu(z,\Omega_{\rm m}, \Omega_{\rm DE},w)$. A negative residual indicates a SN that is too bright (post light-curve correction) for its luminosity distance. Our binned error-weighted mean residual is plotted as red squares and individual measurements are grey points. The bins are constructed such that each bin is of equal width with the point centred on the midpoint of the bin. The error bar along the abscissa extends across the length of the bin. The error bar along the ordinate is the standard error in the weighted mean.

We do not see any trend between $x_{1}$ and Hubble residual, indicating that our single $\alpha$ value represents the entire range of $x_{1}$ values. A slight deviation is seen in our final $x_{1}$ bin, but it is not statistically significant ($< 2 \sigma$).

In the bottom panel, we do find a linear trend with $c$ and the Hubble residual.  SNe with $c <-0.05$ have a negative slope with respect to Hubble residual, implying that our best-fitting $\beta$ value is not applying an appropriate linear correction in this range of SN colour. This trend corresponds  to a preference for a smaller $\beta$ for low $c$ values. We find a slight deviation in our final $c$ bin, but it is not statistically significant ($< 2 \sigma$).

\cite{sullivan11b} found a similar trend between $c$ and Hubble residual in their analysis of SNLS3 data. The authors attribute the trend in Hubble residual to a dependence of $\beta$ on the stellar mass of the galaxy. When splitting their sample based on host-galaxy mass, they found strong evidence (\about\,4.5$\sigma$) that low-mass galaxies prefer a higher $\beta$ value than high-mass galaxies. Since many of our bluest, least reddened SNe are found preferentially in early-type (and presumably massive) galaxies (see Section \ref{ss:host_gal}), this may explain the trend we see in Figure \ref{figure:galmorph}. A similar result is found by \cite{lampeitl10a} in a study of the SDSS sample (\about\,3.5$\sigma$). In a future paper, we will further explore trends between our sample and host-galaxy properties.

Alternatively, the different $\beta$ values may be the result of two different SN~Ia populations that prefer different $\beta$ values. \cite{wang09a} and \cite{silverman12c} found that objects with high-velocity (HV) \ion{Si}{II} $\lambda 6355$ features in their spectra near maximum brightness have an observed distribution of $(B - V)_{\rm max}$ colours that is redder compared to that of spectroscopically normal objects. \cite{wang09a} found that when separating SNe into two classes by \ion{Si}{II} velocity, the two classes preferred different reddening laws (i.e., different $\beta$ values). The HV objects with a redder colour distribution preferred a smaller beta compared to normal objects. This runs counter to what we observe in our data, where the bluest objects appear to prefer a smaller $\beta$. We note, however, that making a cut on $c$ differs from making a cut by spectroscopic subclass since the two observed colour distributions are offset by only \about\,0.10 mag and overlap considerably. Future observing programs that include spectroscopic follow-up observations may be able to disentangle these effects. We explore this further in Section \ref{ss:vel} using a subset of the low-$z$ sample that also has spectral measurements.

\subsection{Systematic Uncertainties}\label{ss:sys_unc}
Thus far, we have only treated the statistical error associated with our SN fits. In this section, we investigate the important role of systematic uncertainties. We consider a systematic error to be an error that affects multiple SNe in our analysis, likely in a correlated manner. In a recent analysis by the SNLS team, \cite{conley11a} found that the most significant contribution to their systematics was calibration uncertainties. We focus our attention on understanding how the calibration affects our results.

A major roadblock to improvement in measurements of cosmological parameters is the inability to accurately measure the zero-point of a photometric system. \cite{ganeshalingam10a} found that the differences between SNe in common between the LOSS and CfA3 samples could be as large as \about\,0.1 mag (in the Landolt photometry system), although the mean difference over all SNe in common is consistent with 0.0 mag. Similarly, in a comparison of the LOSS and CSP photometry for 14 SNe in common, \cite{ganeshalingam11a} found differences of $0.03$~mag in the $B$ and $V$ bands. These comparisons did not include S-corrections to place the SNe on a common photometric system.

As a measure of systematic errors between data sets, we compare SALT2 light-curve parameters for overlapping SNe in Table \ref{t:lcpc}. SALT2 measures parameters in SED space using transmission curves for each photometry system which should alleviate any concerns about S-corrections. We find an excellent level of agreement among the 11 overlapping objects in the LOSS and and CSP data sets for measurements of $m_{B}$ and $c$. There are indications of a \about\,0.02 mag offset between LOSS and the other two samples at significant levels (2.4$\sigma$ for Lit and 4$\sigma$ for CfA3). 

Comparing the LOSS $x_{1}$ values with the CfA3 and CSP samples, we find differences of \about\,0.1, which translate to differences of \about\,0.01 mag for typical $\alpha$ values (0.12--0.15). LOSS $x_{1}$ values best agree with the Lit sample. All of our samples have the same $c$ values to within $1\sigma$. We do not see a significant trend of the differences between LOSS and another sample with time which would indicate an issue with one of the KAIT filter/CCD combinations (KAIT[1--4]).

A similar comparison between nine objects in both the CfA3 and CSP samples shows somewhat large offsets in $m_{B}$ of $0.054 \pm 0.012$ mag and in $x_{1}$ of $0.115 \pm 0.032$ (error-weighted mean and uncertainty). The $c$ values of the two samples agree to within uncertainties. 

\small
\begin{table}
\caption{SALT2 light-curve parameter comparison \label{t:lcpc}}
\begin{center}
\begin{tabular}{l  c | c   c c c   c}
\hline
& & Mean & Std. Dev & WMean ($\sigma_{\rm WM}$) & $N$ \\
\hline
\multicolumn{2}{l}{} &
\multicolumn{4}{c}{$m_{B}$ (mag)} \\ \hline
LOSS $-$ LIT && $-$0.021 &  0.039 & $-$0.024 (0.010)&18 \\
LOSS $-$ CFA3&  & $-$0.024 &  0.058 & $-$0.024 (0.006)&47 \\
LOSS $-$ CSP & & 0.001 &  0.037 & 0.002 (0.011)& 11\\
CFA3 $-$ CSP  && 0.054 &  0.037 & 0.054 (0.012)& 9\\
& & & & & \\ \hline
\multicolumn{2}{l}{} &
\multicolumn{4}{c}{$x_{1}$} \\ \hline
LOSS $-$ LIT&& $-$0.048 &  0.261 & $-$0.009 (0.026) & 18\\
 LOSS $-$ CFA3& & $-$0.082 &  0.328 & $-$0.083 (0.019)& 47\\
LOSS $-$ CSP & &0.048 &  0.269 & 0.124 (0.030)& 11\\
CFA3 $-$ CSP& &0.074 &  0.201 & 0.115 (0.032)& 9\\
& & & & & \\ \hline
\multicolumn{2}{l}{} &
\multicolumn{4}{c}{$c$} \\ \hline
LOSS $-$ LIT& & $-$0.008 &  0.034 & $-$0.020 (0.026) & 18 \\
LOSS $-$ CFA3&  & $-$0.016 &  0.032 & $-$0.019 (0.019) & 47 \\
LOSS $-$ CSP & & $-$0.004 &  0.032 & $-$0.005 (0.030) & 11 \\
CFA3 $-$ CSP& &0.011 &  0.025 & 0.015 (0.032) & 9 \\
\hline 
\end{tabular}
\end{center}
\end{table}
\normalsize

We have also performed SALT2 fits for the LOSS photometry on the Landolt system (without S-corrections) and find overall that the results agree well. We do note, however, that we find an error-weighted mean difference of $0.014 \pm 0.004$~mag in $m_{B}$. Differences in $c$ and $x_{1}$ are consistent with no difference. Comparing our fits in the standard system to the CfA and Lit improves the offset to \about\,0.01 mag, but worsens the comparison to CSP to \about0.04 mag. The weighted mean residuals for $x_{1}$ and $c$ mostly remain the same.

The level of disagreement between the different data sets is somewhat alarming, but it is also within the systematic uncertainties attributed to the photometric zero-points associated with each sample. However, when combined with the other higher-redshift samples, the cumulative effect can alter best-fitting cosmological parameters and the associated uncertainties. 

To model the effect of these uncertainties, we perform a Monte Carlo simulation focusing on the effects of the zero-point uncertainty for the low-$z$ samples. For each filter in each photometry sample, we draw a random systematic uncertainty to add to the SN photometry in that sample's filter. The uncertainty is drawn from a Gaussian distribution centred on 0.0 mag with a $\sigma$ of 0.02 mag. Our value for $\sigma$ is based on the mean offsets between different photometric samples which hint at a zero-point uncertainty of \about\,0.02--0.03 mag. A realization of a dataset includes perturbed photometry for all of the photometry samples. Each dataset realization is fit by SALT2 for new light-curve parameter estimates and then run through our $\chi^{2}$ minimisation code to find the best-fitting cosmological parameters.

When including the zero-point uncertainty assuming a flat Universe, we find that our systematic error translates into an additional systematic error $\Delta w = 0.116$ and $\Delta \Omega_{\rm m} = 0.05$.  Relaxing our assumption on flatness, but requiring DE described by a cosmological constant, we find $\Delta \Omega_{\rm m} = 0.039$ and $\Delta \Omega_{\rm DE} = 0.098$.

While we have focused only on the calibration zeropoint of the low-$z$ photometry systems, an exhaustive treatment of systematic errors by \cite{conley11a} found a similar level of systematic uncertainty in their cosmological parameter estimates. However, they found only $\Delta w = 0.065$ from the uncertainty of the low-$z$ zero-points. They attributed a zero-point uncertainty of \about\,0.01 mag for the various low-$z$ samples which we believe is likely an underestimate. \cite{hicken09b} quoted a systematic uncertainty of $ \Delta w = 0.11$, but this value includes combined constraints from a BAO prior from \cite{eisenstein05a} which decreases the systematic error from SNe alone. \cite{bernstein12a} provide an estimate for the systematic errors for SNe expected to be found as part of DES.


\section{Physical Systematics in the Low-Redshift Sample} \label{s:phy_sys}
In this section, we focus exclusively on looking for physical systematics inferred from the low-$z$ samples. The low-$z$ samples have the advantage of often having complementary spectroscopic coverage and publicly available data on host galaxies. We fix our cosmological parameters to concordance cosmology values of $(\Omega_{\rm m}, \Omega_{\rm DE},w) = (0.27,0.73,-1)$ and our fitting coefficients $(\alpha,\beta) = (0.146,3.169)$. We note that these values of $(\alpha,\beta)$ are consistent with the values found if we fix $(\Omega_{\rm m}, \Omega_{\rm DE},w) = (0.27,0.73,-1)$ and fit for the coefficients using only the low-$z$ sample; we find $\alpha = 0.154 \pm 0.011$ and $\beta = 3.132^{+0.113}_{-0.111}$.

\subsection{Is There a Hubble Bubble?} \label{ss:hubble_bubble}
\cite{zehavi98a} and \cite{jha07a} presented evidence for a monopole in the peculiar velocity field of galaxies in the local Universe. Their analyses indicated that we live in an underdense region of the Universe relative to the global average energy density, manifesting itself as a larger Hubble constant within $cz < 7500$~\kms. Both analyses adopted distances using the MLCS framework and found the presence of a ``Hubble bubble" at \about\,$2\sigma$ confidence assuming a flat, $\Lambda$-dominated Universe with $\Omega_{\rm m} = 0.3$. \cite{zehavi98a} used a previous version of MLCS with a smaller training set compared to the version (MLCS2k2) used by \cite {jha07a}. In a reanalysis of the data from \cite{jha07a}, \cite{conley07a} found that the presence of the Hubble bubble may actually be an artefact of assuming that dust in other galaxies has the same extinction properties as Milky Way dust with $R_{V} = 3.1$ ($\beta = 4.1$).  \cite{conley07a} found that the Hubble bubble disappeared using lower values of $R_{V}$ as preferred by the data when doing a $\chi^2$ minimisation of SALT2 parameters similar to the analysis presented here. Our best-fitting $\beta = 3.169$ corresponds to $R_{V} = 2.169$ under the assumption that the SALT2 $c$ is an estimate of $E(B-V)$ host-galaxy reddening.

\cite{hicken09b} found that the significance and partition redshift of the Hubble bubble was a function of light-curve fitter, assumption of host-galaxy extinction, and even photometry sample.  They analysed departures from a single Hubble law using low-$z$ SN distances estimated from SALT, SALT2, MLCS2k2 with $R_{V} = 3.1$, and MLCS2k2 with $R_{V} = 1.7$. \citeauthor{hicken09b} found the most significant detection of a Hubble bubble at the \about\,$5 \sigma$ confidence level at  $cz = 8400$~\kms\ with MLCS2k2 and $R_{V} = 3.1$. The significance of the void decreased substantially to 1.3$\sigma$ confidence if SNe with $A_{V} > 0.5$~mag are excluded. \citeauthor{hicken09b} did not find evidence of a Hubble bubble using distances derived from SALT or SALT2.

\begin{figure}
\begin{center}
\includegraphics[scale = .55]{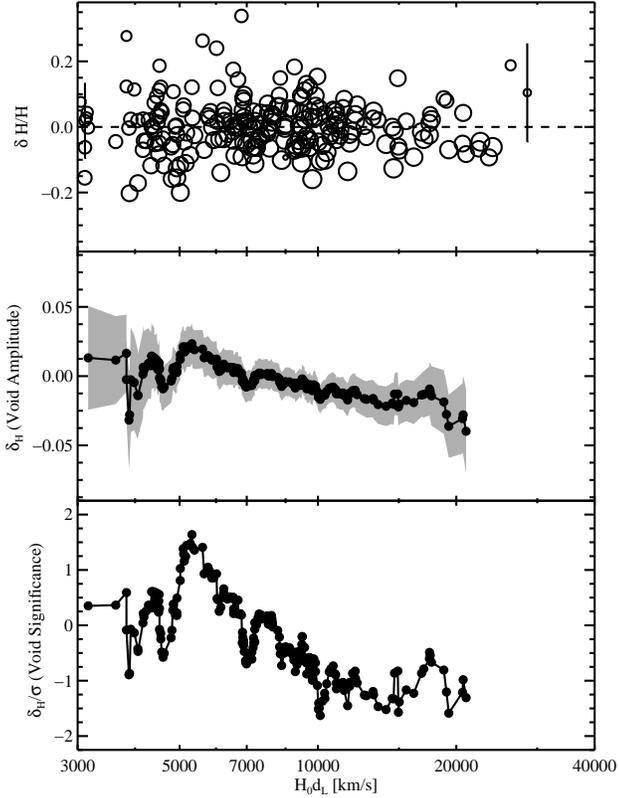}
\end{center}
{\caption[Significance of a Hubble bubble]{Significance of a Hubble bubble in our low-$z$ data. The top panel shows the fractional deviation from the distance luminosity for each object in the low-$z$ sample. The size of each point is inversely proportional to its uncertainty. In the middle panel, we plot the amplitude of a Hubble void signature as a function of redshift. In the bottom panel, we plot the significance of the void amplitude as a function of amplitude. We do not find a significant detection of a Hubble bubble in our data.}\label{figure:hubble_bubble}}
\end{figure}

Following previous analyses, we work in the more convenient units of \kms\ for distances. SN distance moduli, $\mu_{\rm SN}$, are translated by converting to distance (Mpc) and multiplying by H$_{0}$. The luminosity distance to $z_{\rm CMB}$ in units of \kms\ for a given cosmology is 
\begin{equation}
{\rm H}_{0}d_{L}(z_{\rm CMB}) = {\rm c}(1+z_{\rm CMB}) \int^{z_{\rm CMB}}_{0}\frac{dz^{'}}{[\Omega_{\rm m}(1+z^{'})^3 + \Omega_{\Lambda}]^{1/2}}.
\end{equation}
The peculiar velocity of a galaxy, $u$, with respect to the CMB rest frame is given by $u = {\rm H}_{0} d_{L}(z_{\rm CMB}) - {\rm H}_{0}d_{\rm SN}$. We note that c in the above equation refers to the speed of light. The fractional deviation from a Hubble flow is $\delta {\rm H}/{\rm H} = u/{\rm H}_{0}d_{\rm SN}$. We plot $\delta {\rm H}/{\rm H}$ for our low-$z$ sample in the top panel of Figure \ref{figure:hubble_bubble}. For the most part, deviations from Hubble expansion appear randomly distributed, consistent with a single value of H$_{0}$.

We estimate the void amplitude signal by partitioning our low-$z$ sample into two subsamples based on redshift. Starting at $z_{\rm partition} =0.0106$, we compute H$_{0}$ in the inner subsample ($z \leq z_{\rm partition}$) and in the outer subsample. We define the void amplitude as $\delta_{\rm H} = ({\rm H}_{\rm inner} - {\rm H}_{\rm outer})/ {\rm H}_{\rm outer}$. We calculate the void amplitude as a function of $z_{\rm partition}$ with the constraint that the minimum  number of points in a subsample is $\geq $ 6. We estimate the void significance by weighting the void amplitude by its uncertainty given by Equation 3 of \cite{zehavi98a}. We plot the void amplitude significance in the middle and lower panels of Figure \ref{figure:hubble_bubble}, respectively.

We do not see a significant void amplitude in the entire sample of low-$z$ SNe. The largest void amplitude of $-0.04$ is detected at c$z_{\rm CMB} = 21,000$~\kms, although the significance of this detection is relatively low and likely caused by a small number of objects in our outer subsample. We do not find a significance larger than $1.64 \sigma$, indicating that our data are consistent with a single value of H$_0$. Our maximum void detection occurs at c$z_{\rm partition} = 5300$~\kms.

We further investigate whether we detect a Hubble bubble in each of our low-$z$ data sets. The LOSS dataset has a peak void amplitude of $1.28\sigma$ at c$z \approx 6000$~\kms, which is marginally consistent with what is found for the entire sample. However, the CfA3 and CSP samples show peak void signatures in the opposite direction and in the region c$z \approx 3900$--4200~\kms, although also at a low significance ($< 1.8 \sigma$). The most significant void detection is from the Lit sample with a void significance of $2.4\sigma$ at c$z = 4200$ \kms, but in the opposite direction of the CfA3 and CSP samples. A list of our results is given in Table \ref{t:hubble_bubble}.

\begin{table}
\caption{Peak Hubble bubble significance \label{t:hubble_bubble}}
\begin{center}
\begin{tabular}{lccc}
\hline
Sample & Partition & Void  & Void  \\
 & Redshift (\kms) & Amplitude  & Significance  \\
\hline
Low-$z$     & 5306 &  0.023 & 1.638  \\ 
LOSS     & 6025 &  0.022 & 1.281  \\ 
CfA3     & 4406 & $-$0.040 & 1.757  \\ 
CSP      & 4047 & $-$0.038 & 1.600  \\ 
Lit      & 4167 &  0.058 & 2.408  \\ 
\hline
\end{tabular}
\end{center}
\end{table}

To summarize, in agreement with \cite{hicken09b}, we find that our analysis is consistent with no Hubble bubble. Our measurement of a void signal depends on the photometry sample, but for each sample the detection is mostly at low significance. Our sample excluded extremely reddened objects which likely have different reddening properties than normal SNe~Ia. We find it likely that previous detections of an apparent Hubble bubble were due to the way in which reddening was treated, consistent with the conclusions of \cite{conley07a} and \cite{hicken09b}.

\subsection{Host Galaxies} \label{ss:host_gal}
With our large sample of low-$z$ SNe, we can look for trends as a function of host-galaxy properties. Previous studies comparing Hubble residuals to host-galaxy properties have found that after correcting the SN luminosity for light-curve width and SN colour, SNe in more massive galaxies appear brighter than SNe in their less massive counterparts \citep{kelly10a,sullivan10a,lampeitl10a}. The physical interpretation of this result remains elusive, but it could be caused by a dependence on the progenitor metallicity which would be a function of host-galaxy mass \citep{kasen09a}. In this section, we study our Hubble residuals and light-curve parameters as a function of galaxy morphology (as a proxy for galaxy mass) and projected galactocentric distance (PGCD) to look for trends in our data. Again, we restrict our analysis to low-$z$ objects where discovery information regarding galaxy morphology and offsets from the host-galaxy nucleus are easily available in the International Astronomical Union Circulars (IAUCs) and the  NASA/IPAC Extragalactic  Database\footnote{http://nedwww.ipac.caltech.edu/. }. In cases where host-galaxy offsets are not available, we calculate the offset from the difference between the host-galaxy coordinates and the SN coordinates; examples include objects found by the SNFactory \citep{aldering02a}.

We caution that the low-$z$ sample suffers from a bias to more massive galaxies because of targeted surveys such as LOSS \citep{filippenko01a,li03b,leaman11a}, which account for a vast majority of the discoveries in the nearby Universe after 1998. LOSS preferentially monitors more-massive galaxies to maximize the number of discovered SNe \citep {leaman11a,li11a,li11b}. The low-$z$ set is not a complete sample, and it is likely not an accurate representation of the true host-galaxy demographics. We continue with that caveat in mind.

\begin{table}
\caption{Residual as a function of Galaxy morphology}{\label{t:galmorph}}
\begin{center}
\begin{tabular}{lccccc}
\hline
Morphology & Mean & Std. Dev & WMean & $\sigma_{\rm WM}$ & $N$ \\
\hline
E     & $-$0.017 & 0.196 & $-$0.034 & 0.040 &  21\\
E/S0  & $-$0.021 & 0.152 & $-$0.004 & 0.077 &   5\\
S0    &  0.037 & 0.140 &  0.029 & 0.037 &  23\\
S0a   &  0.113 & 0.268 &  0.097 & 0.081 &   5\\
Sa    &  0.031 & 0.135 &  0.033 & 0.048 &  13\\
Sab   & $-$0.020 & 0.129 & $-$0.004 & 0.054 &  12\\
Sb    & $-$0.032 & 0.176 & $-$0.027 & 0.023 &  57\\
Sbc   & $-$0.066 & 0.181 & $-$0.070 & 0.046 &  14\\
Sc    &  0.022 & 0.239 &  0.017 & 0.035 &  26\\
Scd   &  0.001 & 0.141 &  0.023 & 0.063 &   8\\
Sd/Ir &  0.063 & 0.192 &  0.065 & 0.049 &  11\\ \hline
E--S0       &  0.008 & 0.167 & $-$0.001 & 0.025 &  49\\
S0a--Sc     & $-$0.011 & 0.189 & $-$0.010 & 0.016 & 127\\
Scd/Sd/Irr &  0.037 & 0.171 &  0.049 & 0.039 &  19\\
\hline
\end{tabular}
\end{center}
\end{table}

\subsubsection{Morphology}
\begin{figure}
\begin{center}
\includegraphics[scale = .55]{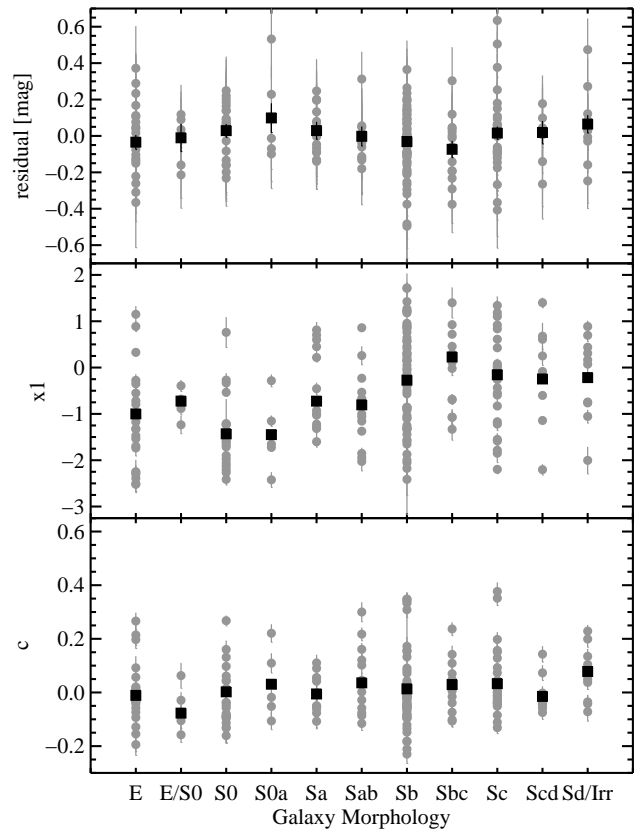}
\end{center}
{\caption[Parameters versus galaxy morphology]{Parameters versus  galaxy morphology. In the top panel we plot the Hubble residual versus host-galaxy morphology. Residuals are measured in the sense that ${\rm residual} = m_{B} - m_{B,{\rm corr}}$. In the middle panel we plot $x_{1}$ versus host-galaxy morphology. In the bottom panel we plot $c$ versus host-galaxy morphology. The red squares are the error-weighted mean value for each morphology bin. We do not find significant evidence of a trend between Hubble residual and host-galaxy morphology.}\label{figure:galmorph}}
\end{figure}

In the top panel of Figure \ref{figure:galmorph}, we plot Hubble residual as a function of host-galaxy morphology. If more-massive galaxies host brighter SNe (after correcting for light-curve width and colour), the expectation is that early-type galaxies (E--S0) should have a more negative mean residual in comparison to late-type galaxies (Scd/Sd/Irr). We find a difference in the error-weighted mean residuals (red squares) between Sd/Irr and E of $0.099 \pm 0.063$ mag in the sense that E galaxies host brighter SNe post-correction. While this agrees with previous results, our result is significant only at the \about\,1.6$\sigma$ confidence level.

If we bin the data into early (E--S0), mid (S0a--Sbc), and late (Scd/Sd/Irr) galaxy bins, we find a difference between the late and early galaxy bins of $0.050 \pm 0.046$ mag, consistent with no difference. We find a more significant difference between mid and late galaxies of $0.059 \pm 0.029$ mag. The consistency between our early and mid galaxy bins indicates that there may be more to gain by either excluding late galaxies or treating them separately. We caution that there are only 19 objects in our late bin, and the rather large weighted mean residual may be a result of small-number statistics. The full set of Hubble residuals as a function of host-galaxy morphology can be found in Table \ref{t:galmorph}.

\cite{sullivan03a} found that the Hubble residual scatter was minimized by using only SNe in early-type hosts, while \cite{hicken09a} found that the scatter was reduced by using late-type galaxies. Our analysis finds that both of these galaxy bins produce a scatter that is slightly lower compared to that of mid galaxies. Early galaxies do slightly better ($\sigma = 0.167$~mag) than late galaxies ($\sigma = 0.171$~mag), although the difference is statistically insignificant. mid galaxies have $\sigma = 0.189$~mag, but they also have twice the number of objects in the early and late galaxy bins combined.

In the middle panel of Figure \ref{figure:galmorph}, we plot the SALT2 $x_{1}$ distribution as a function of host-galaxy morphology. SNe in early-type galaxies have a smaller weighted $x_{1}$ average (narrower light curves, corresponding to underluminous SNe) compared to SNe in mid- and late-type galaxies. This observation has been noted in previous studies \citep{della-valle94a,hamuy96c,howell01a,hicken09b}.

The presence of SNe~Ia in both passive elliptical galaxies with old stellar populations and in spirals with active star formation has led to speculation that SNe Ia may come from at least two progenitor populations \citep{mannucci05a,scannapieco05a, sullivan06a,neill06a,maoz11a}.  SNe~Ia in late-type galaxies are typically brighter and occur at \about\,10 times the rate (per unit mass) of SNe~Ia in elliptical galaxies, leading to the suggestion that these SNe may be linked to young stellar progenitors. SNe~Ia in early-type galaxies, on the other hand, more likely track the total mass of the galaxy. This has led rate studies to adopt a two-component model: a ``prompt" component proportional to the star-formation rate and a ``tardy" component proportional to the total mass. Recently, in an analysis of SN rates from LOSS, \cite{li11b} found intriguing evidence of a rate-size relation across all galaxy classes, indicating that the SN~Ia rate is {\it not} linearly proportional to galaxy mass. Specifically, the SN rate per unit mass is larger in smaller galaxies compared to larger galaxies. Our analysis hints that different galaxy morphologies host different distributions of SN properties. However, we again caution that our distributions are likely biased and may not reflect the true distribution of SN properties. 

In the bottom panel of Figure \ref{figure:galmorph}, we plot the SALT2 $c$ distribution as a function of galaxy morphology. The reddest SNe occur in Sd/Irr galaxies and the bluest in E--S0. This result is somewhat counterintuitive given the above result that early galaxies host underluminous SNe that are intrinsically redder \citep{riess96b,jha07a}. This may instead reflect the distribution of host-galaxy extinction values. Early-type galaxies are expected to have a lower dust content and minimal amounts of star formation compared to late-type galaxies which generally exhibit more star formation. The bluer colours in early-type galaxies may reflect the lack of host-galaxy extinction. SNe in our middle galaxy bins show a larger distribution of $c$ values.

\subsubsection{Projected Galactocentric Distance}
\begin{figure}
\begin{center}
\includegraphics[scale = .55]{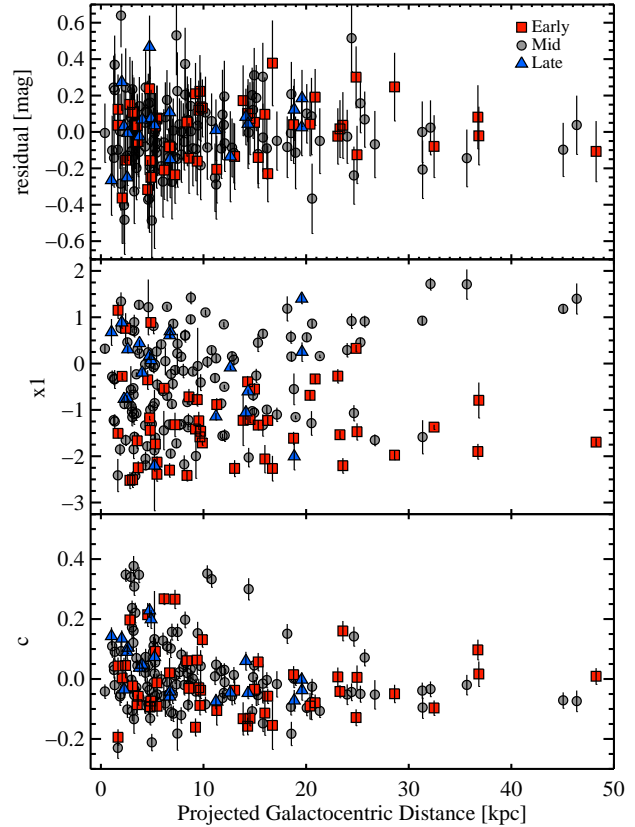}
\end{center}
{\caption[Parameters versus projected galactocentric distance]{Parameters versus projected galactocentric distance (PGCD). The top panel shows Hubble residual versus PGCD. The middle panel is $x_{1}$ versus PGCD, and the bottom panel is $c$ versus PGCD. Points are colour coded by galaxy type. We find that the reddest SNe occur within 15 kpc of the host-galaxy nucleus.}\label{figure:res_v_gcd}}
\end{figure}

Similar to the above discussion of host-galaxy morphology, we now turn to an analysis with projected galactocentric distance (PGCD). In Figure \ref{figure:res_v_gcd}, we plot Hubble residual (top panel), $x_{1}$ (middle panel), and $c$ (bottom panel) as a function of PGCD. Points are colour coded by their host-galaxy morphology as either early (red squares), mid (grey circles), or late (blue triangles) galaxies using the definition of the previous subsection.

We do not see any significant trends of Hubble residual with PGCD or with host-galaxy classification. There is a noticeable lack of SNe in late-type galaxies at a PGCD $> 20$ kpc which was also noticed by \cite{hicken09b}. Partitioning our sample into two subsamples using a cut of 10 kpc in PGCD, we find that the inner sample has $\sigma = 0.186$~mag and the outer sample has $\sigma = 0.156$~mag. The larger Hubble residual from SNe within 10 kpc may be the result of systematic uncertainties due to the difficulty in galaxy template subtraction or the increased effects of host-galaxy extinction. Future SN~Ia samples may be able to reduce the scatter in Hubble diagrams by only using SNe at a PGCD $> 10$ kpc.

In the middle panel of Figure \ref{figure:res_v_gcd}, we plot $x_{1}$ against PGCD. As mentioned in the previous subsection, SNe~Ia are primarily underluminous (small $x_{1}$) in early-type galaxies  and overluminous (larger $x_{1}$) in late-type galaxies. Mid-type galaxies show a rather uniform distribution within the inner 20 kpc of the host. However, at distances $ > 20$~kpc, the error-weighted $x_{1}$ mean increases from $-0.473 \pm 0.092$ to $0.345 \pm 0.284$ (standard error of the mean), corresponding to a 2.7$\sigma$ difference.

In the bottom panel of Figure \ref{figure:res_v_gcd}, we plot $c$ versus PGCD. As one might naively expect, the reddest SNe are located primarily within the inner 10\,kpc where host-galaxy reddening will likely affect SN colour.  Moving out to larger PGCD, SN colour becomes increasing bluer for all of our galaxy classes.

\subsection{Silicon Velocity}\label{ss:vel}
Recent attention has been drawn to incorporating velocity information into distance fits. \cite{wang09a} found that separating SNe~Ia into normal and high-velocity (HV) classifications based on  \ion{Si}{II} $\lambda$6355 velocity near maximum light significantly improved distance estimates.  \cite{wang09a} found that the two classes have different observed $B_{\rm max} - V_{\rm max}$ distributions while having similar $\dmb$ distributions. Performing a $\chi^2$ minimisation fit in the same vein as Equation \ref{eqn:chi2}, the authors found that the HV SNe preferred a lower $\beta$ than the normal SNe, indicating either a difference in the intrinsic colour distribution and/or differences in the reddening law for the two samples.

In this section, we look for trends between our derived light-curve parameters, host-galaxies properties, and \ion{Si}{II} velocity as measured by the minimum of the blueshifted feature of the $\lambda$6355 line within 5 days of maximum light. Previous studies by \cite{blondin11a} and \cite{silverman12c} found that including velocity information along with light-curve parameters did not substantially improve distance estimates.  We match objects in our low-$z$ photometry sample presented here to the Berkeley Supernova Ia Program (BSNIP) spectroscopic sample \citep{silverman12a}. We also apply the software and algorithmic framework laid out by \cite{silverman12b} to measure \ion{Si}{II} velocities in the near-maximum-light spectra published by \cite{blondin11a}. We have photometric and spectroscopic measurements for 65 objects. We note that LOSS light-curve parameters measured using light curves in the Landolt system were used in a similar study by \cite{silverman12c}. Here, we use the light-curve parameters derived from the LOSS natural-photometry system.

Building upon this work, \cite{foley11a} used the \cite{wang09a} sample and attributed the difference in $\beta$ to differences in the intrinsic reddening distribution of the two subclasses. Using the CfA spectroscopy sample, \cite{foley11b} found that splitting objects into these two classes based on \ion{Si}{II} velocity at maximum light improves distance measurements.

\begin{figure}
\begin{center}
\includegraphics[scale=.55]{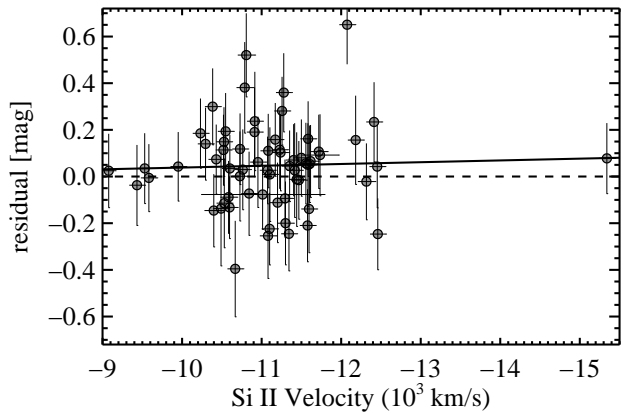}
\end{center}
{\caption[Hubble residual versus Si~II velocity]{Hubble residual versus \ion{Si}{II} velocity. Overplotted as a solid line is a linear fit to the data. The slope is consistent with 0, indicating no significant correlation between Hubble residual and Si~II velocity.}\label{figure:vel_v_res}}
\end{figure}

Rather than separating SNe into two classes, we seek to assess whether incorporating $v_{\rm Si}$ as a continuous parameter along with light-curve parameters improves distance estimates. In Figure \ref{figure:vel_v_res}, we plot the Hubble residual versus the measured \ion{Si}{II} $\lambda 6355$ velocity for 65 objects in our low-$z$ sample having spectral measurements from BSNIP. There is no obvious trend between residual and velocity. Performing a Bayesian Monte Carlo linear regression \citep{kelly07a}, we find a slope consistent with 0, indicating no statistically significant correlation between the two variables (dashed line). We find a Pearson linear-correlation coefficient of $-0.044 \pm 0.084$, consistent with no correlation.

\cite{foley11a} found that distance estimates to SNe~Ia may be improved by taking advantage of a relationship between \ion{Si}{II} velocity at maximum light and ``intrinsic colour" \citep{foley11b,foley12a}. \cite{foley11b} reported an $3.4\sigma$ relationship between ``intrinsic colour" and \ion{Si}{II} velocity that, when accounted for, decreases the dispersion in distance estimates. However, their definition of intrinsic colour assumed a somewhat simplistic offset from the observed $B_{\rm max} - V_{\rm max}$ pseudocolour. \cite{blondin12a} performed a more sophisticated measurement of intrinsic $B_{\rm max} - V_{\rm max}$ colour with rigorous handling of uncertainties and found a significantly reduced correlation at the $2\sigma$ level. Their intrinsic colour estimates were derived from  BayeSN \citep{mandel09a,mandel11a}, which incorporates population correlations between intrinsic absolute magnitudes, intrinsic colours, light-curve shape, and host-galaxy properties. While our work makes use of the SALT2 $c$ parameter which is more closely related to the $B_{\rm max} -V_{\rm max}$ observed colour, we do not see evidence that including a linear correction for \ion{Si}{II}  velocity improves distance estimates.

\begin{figure}
\begin{center}
\includegraphics[scale=.55]{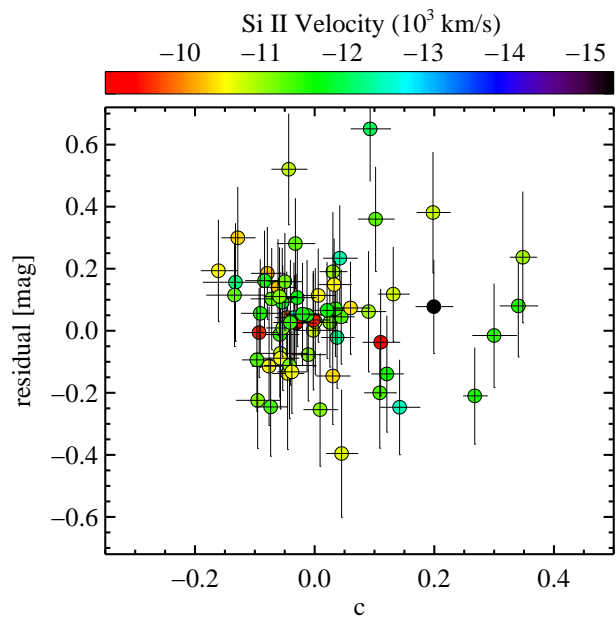}
\end{center}
{\caption[Hubble residual versus $c$, colour coded by velocity]{Hubble residual versus $c$ colour coded by Si~II velocity. We do not see an obvious trend with respect to where higher velocity objects are located in this plot.}\label{figure:res_v_c_w_v}}
\end{figure}

In Figure \ref{figure:res_v_c_w_v}, we plot the Hubble residual versus $c$ (similar to the bottom panel of Figure \ref{figure:param_v_res}) colour coded by \ion{Si}{II} velocity. In Section \ref{ss:res} we noted a trend between Hubble residual and $c$ indicating that the bluest objects preferred a smaller $\beta$ value. The linear trend is still visible with fewer objects (65 here versus 586 in Section \ref{ss:res}). \cite{wang09a} and \cite{foley11a} find that objects classified as high velocity ($v_{\rm Si~II} >  11,800$ \kms) have a redder colour distribution compared to normal objects. This is not evident in our plot, indicating that the linear trend between $c$ and Hubble residual does not correspond to a correlation with $v_{\rm Si~II}$.

\begin{figure}
\begin{center}
\includegraphics[scale=.55]{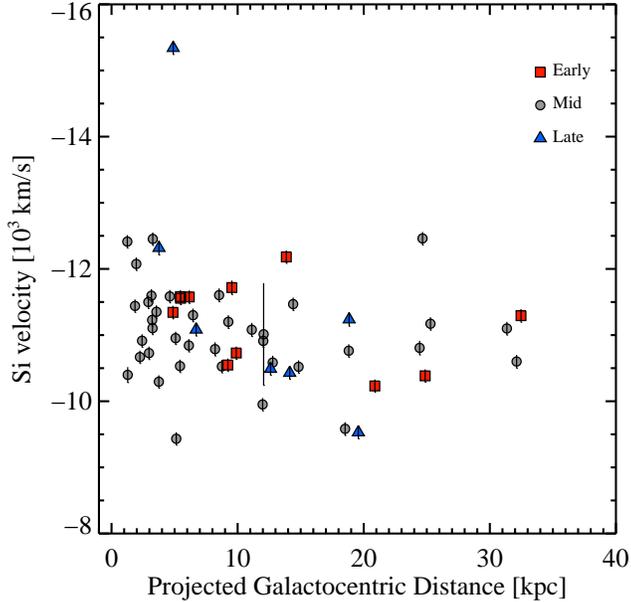}
\end{center}
{\caption[Si~II velocity versus projected galactocentric distance]{Si~II velocity versus projected galactocentric distance. Objects are colour-coded by galaxy type. Early-type galaxies (E--S0) are red squares, mid-type galaxies (S0a--Sbc) are grey circles, and late-type galaxies (Scd/Sd/Irr) are blue triangles. We do not see a significant difference between velocity and galaxy type, although the average Si~II velocity appears to slightly decrease with PGCD.}\label{vel_v_gcd}}
\end{figure}

In Figure \ref{vel_v_gcd}, we plot the \ion{Si}{II} velocity for 56 objects having spectroscopic and host-galaxy information. We use the same colour code as in the bottom panel of Figure \ref{figure:res_v_gcd}, with red squares representing early galaxies (E--S0), grey circles as mid galaxies (S0a--Sc), and blue triangles for late galaxies (Scd/Sd/Irr). We do not see an obvious trend regarding the distribution of \ion{Si}{II} velocities in any of these galaxy bins. Similarly, \cite{wang09a} found that Normal and HV objects reside in similar host-galaxy distributions (see their Figure 3d). Dividing our sample into HV and Normal objects using the criterion of \cite{wang09a}, we find that both objects have a similar PGCD distribution based on a two-sided Kolmogorov-Smirnoff test (68\% probability the two samples are drawn from the same parent distribution). 

For the above analysis, we used SNe with velocities taken within 5 d of maximum light in the $B$ band. We have also performed our analysis using the family of velocity evolution curves presented by \cite{foley11b} derived from the CfA spectral sample \citep{blondin12a} to extrapolate our velocities at various phases to velocity at maximum light in the $B$ band.  The family of curves is derived for SNe with $1.0 < \Delta m_{15}(B)< 1.5$~mag with velocity measurements made within a week of maximum light under the assumption that a \ion{Si}{II} velocity is proportional to its time gradient.  Using this relation does not change any of the results presented here.

Previous work by \cite{wang09a} and \cite{foley11a} found that separating objects into two classes with different reddening laws improved distance estimates. We do not find any evidence that including \ion{Si}{II} velocity as a continuous parameter improves distance estimates or correlates to trends seen in SALT2 $c$. A more detailed analysis that can infer the intrinsic colour of SNe by disentangling contributions from host-galaxy reddening is required to better test how velocities may improve distance estimates.


\section{Conclusions} \label{s:cosmo_conclusions}
We have presented a cosmological analysis of the first LOSS data release of the light curves of 165 SNe~Ia in addition to other photometry data sets available in the literature. Rather than fitting the LOSS photometry in the Landolt system, we have rereduced our data on the natural system of the KAIT and Nickel telescopes and implemented SALT2 with accurate throughput curves for our telescope systems. Both the Landolt and natural-system light curves for our SN~Ia sample as well as the modifications to the SALT2 package is available to the community via the LOSS group page\footnote{http://hercules.berkeley.edu/database .}.

From our SN~Ia measurements alone, we determine $w =-0.86^{+0.13}_{-0.16}\rm{(stat)} \pm 0.11{\rm (sys)}$, consistent with a cosmological constant. We reject a Universe devoid of DE or a failure of general relativity at the $> 4\sigma$ level (99.999\% confidence). When combined with other probes of cosmology such as BAO \citep{percival10a} and measurements of anisotropy in the CMB \citep{komatsu11a}, we find improved constraints of $w = -1.067^{+0.050}_{-0.046}$. Allowing for curvature in a Universe with a cosmological constant, our results are consistent with a flat, $\Lambda$-dominated Universe.

Using a Monte Carlo routine, we investigate the role that zero-point calibration uncertainties play in the error budget for cosmological parameters. Motivated by mean offsets between different low-$z$ data sets for overlapping SNe, we adopt a typical uncertainty of 0.02 mag for the zero-point for each filter bandpass of each photometry system. Our uncertainty in the calibration zero-point from the low-$z$ data sets propagates into an uncertainty of 0.11 in $w$. This is smaller than the statistical uncertainty, but still a significant contribution to the total uncertainty. Future low-$z$ SN~Ia samples will be best served by adopting a natural-photometry system in common with the higher-$z$ samples.

Our data are consistent with no existence of a monopole in the galaxy velocity field produced by a relative underdensity in the local Universe (i.e., a Hubble bubble). Previous detections were likely due to the treatment of SN colour. In this work, we adopt a linear correction term that prefers a coefficient for SN colour which implies $R_{V} \approx 2.2$. Previous detections of the Hubble bubble enforced a prior of $R_{V} = 3.1$ on host-galaxy dust. Our results agree with the analyses of \cite{conley07a} and \cite{hicken09b}.

We find no significant correlations between Hubble residuals and host-galaxy properties. This runs counter to recent studies that find massive galaxies host brighter SNe after correcting for light-curve width and SN colour \citep{kelly10a,sullivan10a,lampeitl10a}. We caution, however, that our work used galaxy morphology as a proxy for galaxy mass. We find that early-type (E--S0) and mid-type (S0a--Sc) galaxies have a similar mean residual, while late-type (Scd/Sd/Irr) galaxies have a larger relative residual. The scatter in Hubble residuals is roughly the same in these three galaxy bins. Future studies by our group incorporating more detailed galaxy properties from stellar population synthesis models will better answer whether our dataset agrees with claimed trends that depend on host-galaxy mass.

We do not find evidence that including the \ion{Si}{II} $\lambda 6355$ velocity near maximum light as a continuous parameter improves distance estimates. This confirms the results of \cite{blondin11a} and \cite{silverman12c}. 

Improvements to measurements of cosmological parameters with SNe Ia rest on our ability to limit systematic uncertainties. Our analysis shows that zero-point calibration uncertainties between the various low-$z$ samples contribute a sizable fraction to the error budget. Future data sets taken with a single telescope system, such as Pan-STARRS, LSST, and DES, will greatly advance our ability to test cosmological models.

\section*{Acknowledgments}
We thank the Lick Observatory staff for their assistance with the operation of KAIT. We are grateful to the many students, postdocs, and other collaborators who have contributed to KAIT and LOSS over the past two decades, and to discussions concerning the results and SNe in general --- especially S. Bradley Cenko, Ryan Chornock, Ryan J. Foley, Saurabh W. Jha, Jesse Leaman, Maryam Modjaz, Dovi Poznanski, Frank J. D. Serduke, Jeffrey M. Silverman, Nathan Smith, Thea Steele, and Xiaofeng Wang. We also thank Alex Conley for many of the cosmology fitting routines used in this work. This research, which took many years, was completed after the untimely and tragic death of our close friend, colleague, and coauthor Weidong Li, who led the nightly operation of KAIT and conducted numerous scientific studies with it; we miss him dearly.

The work of A.V.F.'s supernova group at UC Berkeley has been generously supported by the US National Science Foundation (NSF; most recently through grants AST--0908886 and AST-1211916), the TABASGO Foundation, the Christopher R. Redlich Fund, US Department of Energy SciDAC grant DE-FC02-06ER41453, and US Department of Energy grant DE-FG02-08ER41563.  KAIT and its ongoing operation were made possible by donations from Sun Microsystems, Inc., the Hewlett-Packard Company, AutoScope Corporation, Lick Observatory, the NSF, the University of California, the Sylvia \& Jim Katzman Foundation, the Christopher R. Redlich Fund, the Richard and Rhoda Goldman Fund, and the TABASGO Foundation. We give particular thanks to Russell M. Genet, who made KAIT possible with his initial special gift to A.V.F.; Joseph S. Miller, who allowed KAIT to be placed at Lick Observatory and provided staff support; Jack Borde, who provided invaluable advice regarding the KAIT optics; Richard R. Treffers, KAIT's chief engineer; and the TABASGO Foundation, without which this work would not have been completed. We made use of the NASA/IPAC Extragalactic Database (NED), which is operated by the Jet Propulsion Laboratory, California Institute of Technology, under contract with NASA.

\bibliographystyle{mn2e}

\appendix
\section{LOSS Natural-System Photometry} \label{app:nat_phot}
Deriving SN photometry in the natural system of the telescope requires determining the natural-system magnitudes of local field standards. In \cite{ganeshalingam10a}, we described the process of deriving Landolt magnitudes of local standard stars  on photometric nights during our observing campaign. We transform the Landolt magnitudes to natural-system magnitudes using the set of equations
\begin{eqnarray}\label{e:cc}
b &=& B + C_{B} (B - V) , \label{e:cc1}\\
v &=& V + C_{V} (B - V) , \\
r&=& R + C_{R} (V - R) ,~\rm{and} \\
i &=& I + C_{I} (R - I), \label{e:cc2}
\end{eqnarray}
where lowercase letters present natural-system magnitudes, uppercase letters represent Landolt-system magnitudes, and $C_{X}$ are the observed colour terms from Table 4 of \cite{ganeshalingam10a} derived from observations of \cite{landolt92a,landolt09a} standards on photometric nights. Catalogs for local field stars for our 165 SNe fields can be downloaded from our website\footnote{http://hercules.berkeley.edu/database .}.

There is a concern that we are introducing a systematic error by transforming our Landolt magnitudes for local field stars to the natural system by simply using the averaged colour terms derived from transforming natural-system magnitudes to Landolt magnitudes. In a sense, we are taking natural-system photometry, transforming it to the Landolt system, and then transforming it back to the natural system. 

A more straightforward approach would create a catalog of \cite{landolt92a,landolt09a} standards on the natural system of each telescope system by applying a linear colour correction (e.g., \ref{e:cc1} -- \ref{e:cc2}). Then for each photometric night, solve for the photometric solution using only a zero-point and an airmass term, and apply that solution to the SN fields to get the natural magnitudes of each local field star. The disadvantage of this approach is that  if a field is calibrated using observations with the Lick Nickel telescope, the local field star magnitudes will be on the Nickel photometry system and cannot be used to calibrate KAIT observations.

To ensure that our technique did not introduce a significant systematic error, we tested the two approaches on a few SN fields and found that the differences were $< 0.01$~mag.

Next, we calculate the zero-points of our bandpasses. The zero-point of a photometric system is defined as
\begin{equation}
ZP_{X} = 2.5 \, \log_{10} \int_{0}^{\infty} F_{\lambda} S_{X}(\lambda) \frac{\lambda}{\rm h c} d\lambda + m_{0},
\end{equation}
where $S_{X}$ is the transmission function for the $X$ band, $F_{\lambda}$ is the SED of the standard star, and  $m_{0}$ is the magnitude of the standard in the Landolt system. Vega has been a historical favourite choice for the photometry zero-point, but it has many shortcomings such as a poorly constrained Landolt magnitude, and it is too bright to be imaged on most modern telescopes \citep{conley11a}. The Supernova Legacy Survey (SNLS) adopted BD +17$^\circ$4806, which has a known Landolt magnitude \citep{landolt07a} and a high-quality {\it Hubble Space Telescope} spectral observation \citep{bohlin04a}.

Instead of determining the zero-point using a single star, we derive the zero-point from the catalog of spectrophotometric standard stars presented by \cite{stritzinger05a}. Adopting the transmission curves given by \cite{ganeshalingam10a}, we synthesize instrumental magnitudes using spectra of 100 spectrophotometric standards\footnote{Download from http://www.das.uchile.cl/$\sim$mhamuy/SPECSTDS/.}. We then fit for the zero-point of each band with the equations

\begin{eqnarray}
b &=& B + C_{B} (B - V) + ZP_{B}, \\
v &=& V + C_{V} (B - V) + ZP_{V}, \\
r&=& R + C_{R} (V - R) + ZP_{R},~\rm{and} \\
i &=& I + C_{I} (R - I) + ZP_{I},
\end{eqnarray}
where lowercase letters are the synthesized instrumental magnitudes, uppercase letters represent Landolt-system magnitudes, $C_{X}$ are colour terms from Table 4 of \cite{ganeshalingam10a}, and $ZP_{X}$ are the desired zero-points of the natural-photometry system. The zero-points of the KAIT[1--4] and Nickel bandpasses can be found in Table \ref{t:zp}.

\begin{table}
\caption{Zero-points for LOSS passbands$^\textrm{a}$ \label{t:zp}}
\begin{center}
\begin{tabular}{lcccc}
\hline
Telescope System & $ ZP_{B}$ & $ ZP_{V}$ & $ ZP_{R}$ & $ ZP_{I}$  \\
\hline
KAIT1 & 15.304 & 14.913 & 15.357 &14.686 \\
KAIT2 & 15.361 & 14.914 &14.975 & 14.452 \\
KAIT3 & 15.332 & 14.921 & 15.008 & 14.457 \\
KAIT4 & 15.249 & 14.922 & 14.973 & 14.439 \\
Nickel & 15.224 & 14.828 & 14.930 & 14.703 \\
\hline
\multicolumn{5}{p{3.5in}}{$^\textrm{a}$zero-points are give in magnitudes.}\\
\end{tabular}
\end{center}
\end{table}

\label{lastpage}

\end{document}